\documentstyle[osa,graphicx,subfigure,manuscript]{revtex}

\begin{document}


\newcommand{\VV}[1]{\mbox{\boldmath{$#1$}}}   
\newcommand{\dif}[2]{\frac{\partial #1}
                          {\partial #2}}      %
\newcommand{\dift}[2]{\partial #1 /
		      \partial #2}           
\newcommand{\diff}[2]{\frac{\partial^{2} #1}%
			   {\partial #2^{2}}} 
\newcommand{\Dif}[2]{\frac{\mathrm{d} \/#1}
                    {\mathrm{d} \/#2}}        %
\newcommand{\Dift}[2]{\mathrm{d} \/#1 /
		     \mathrm{d} \/#2}         
\newcommand{\nab}[0]{\vec{\nabla}}            
\newcommand{\Div}[1]{\vec{\nabla}\cdot\/#1}   
\newcommand{\Dim}[1]{\mbox{$\,\mathrm{#1}$}}  
\newcommand{\eqnref}[1]{(\ref{#1})}           
\newcommand{\fref}[1]{\ref{#1},               %
            Seite \pageref{#1}}               %
\newcommand{\argu}[1]{\! \left( #1 \right)}   
\newcommand{\Achtung}[1]{\emph{#1} 
            \marginpar{\Huge\textbf{!}}}      
\newcommand{\code}[1]{\texttt{#1}}            
\newcommand{\file}[1]{\texttt{#1}}            
\newcommand{\class}[1]{\texttt{#1}}           
\newcommand{\funct}[1]{\texttt{#1}}           
\newcommand{\prog}[1]{\texttt{#1}}            
\newcommand{\ds}[0]{\displaystyle}            
\newcommand{\ts}[0]{\textstyle}               

\newcommand{\invh}[1]
   {\newlength{\hei}%
    \newlength{\dep}%
    \settoheight{\hei}{#1}%
    \settodepth{\dep}{#1}%
    \addtolength{\hei}{\dep}%
    \rule[\dep]{0pt}{\hei}%
   }
\newcommand{\invw}[1]
   {\newlength{\wid}%
    \settowidth{\wid}{#1}%
    \rule{\wid}{0pt}%
   }

\newcommand{\aref}[1]%
	{(\/\textit{Fig.~\ref{#1}}\/)} 	      

\newcommand{\bref}[2]%
	{(\/\textit{Fig.~\ref{#1}(#2)}\/)}    
\newcommand{\ind}[1]{_{\rm{#1}}}              

\begin{center}{\bf Shear-Flow Driven Current Filamentation: Two-Dimensional
Magnetohydrodynamic-Simulations}
\end{center}

\begin{center}{C. Konz, H. Wiechen, and H. Lesch\\
Center for Interdisciplinary Plasma Science (CIPS)\\
Institut f{\"u}r Astronomie und Astrophysik \\
der Universit{\"a}t M{\"u}nchen\\
Scheinerstra{\ss}e 1, D-81679 M{\"u}nchen, Germany}
\end{center}

\begin{center}{\bf ABSTRACT}
\end{center}
The process of current filamentation in permanently
externally driven, initially globally ideal plasmas is investigated by
means of two-dimensional Magnetohydrodynamic~(MHD)-simulations. This situation is
typical for astrophysical systems like jets, the interstellar and intergalactic
medium where the dynamics is dominated by external forces. Two different
cases are studied. In one case, the system is ideal permanently and dissipative 
processes are excluded. In the second case, a system with a current 
density dependent resistivity is considered. This resistivity is switched on 
self-consistently in
current filaments and allows for local dissipation due to magnetic
reconnection. Thus one finds tearing of current filaments and, besides, merging
of filaments due to coalescence instabilities. Energy input and dissipation
finally balance each other and the system reaches a state of constant magnetic
energy in time.

\bigskip


\newpage

\bigskip
\noindent {\bf I. INTRODUCTION}

\smallskip

One of the central issues in plasma astrophysics are dissipative processes in
highly collisionless (from a kinetic point of view) or highly ideal (in the
framework of a fluid description) magnetized gases.
\\
A typical feature for a broad variety of cosmic or space plasma configurations
is the continued input of energy into the system due to some external forces. This
is done e.g. by external shear forces, twisting, or compression which result from
gravity, rotations, or winds. Some of the more important mechanisms are shown in
Figure~\ref{fig:intention}. The external forces are unsaturated on the time
scales considered in this paper such that the system is subject to a permanent
shearing, twisting, or compression. As a
consequence, the magnetic field is disturbed on large spatial scales and the
magnetic energy increases. This process continues as long as there are no dissipative channels available
to get rid of the external energy input. This scenario refers to the
dynamics of the Earth's magnetosphere$^{1}$, solar activity$^{2}$, stellar
jets$^{3}$, the center of the Milky Way$^{4}$ or galactic jets$^{5,6}$ for instance.
\\
One of the most important dissipative processes in astrophysical plasmas is
magnetic reconnection. The conversion of magnetic energy into kinetic particle
energy and into heat can be done by reconnection in a very effective way
because reconnection proceeds on time scales which are by a factor of
$N_L^{1/2}$ ($N_L$: Lundquist number) or more faster than diffusion, demanding
for local deviations from ideal conductivity only$^{7}$.
\\
External forces, like induced shear flows, twisting and shearing the magnetic
field in an ideal astrophysical plasma result in strong local magnetic field
gradients corresponding to strong local enhancements of the electric current
density. Local thin current sheets form and the current density becomes more
and more filamentary$^{8}$.
\\
Thus, an external distortion is redistributed down to smaller and smaller
spatial scales. The magnetic shear length decreases by the continuing twisting of the
magnetic field corresponding
to an increase of the local current density on decreasing spatial scales down
to scales where a dissipative channel opens.
\\
Considering magnetic reconnection, this means that the relevant scales
cascade down until sources for sufficient local deviations from the
macroscopic ideal conductivity become available. This might be the scale of
the ion-gyroradius where current driven micro-instabilities can exite
microscopic fluctuations acting as a local effective resistance (an
anomalous resistivity) on macroscopic scales$^{9}$. Depending on the specific
configurations under consideration the cascade of filamentation has even to
scale down to the scale of the electron inertia length. Especially in thin,
hot collisionless plasmas, the inertia of electrons might be the ultimative
dissipative channel allowing for magnetic reconnection$^{10}$.
\\
Thus, independent of which specific dissipative channel may be the relevant
one, magnetic reconnection in an ideal astrophyical plasma demands for the
development of current sheets, i.e. for a filamentation of the current
density.
\\
The formation of thin current sheets and current filamentation has been
subject of both intensive analytical and simulational studies.
\\
Parker$^{8}$ has shown that shearing of the footpoints of a 2-D
magnetic field yields regions of different topological structures seperated by
tangential discontinuities, i.e. thin current sheets. Considering a
one-dimensional Harris sheet, Hahm and Kulsrud$^{11}$ have shown that singular
current densities in the center of the sheet are a general consequence of
widely arbitrary, ideal boundary perturbations. More detailed studies of
current sheet formation in ideal Magnetohydrodynamic~(MHD)-systems can be 
found in e.g. Schindler
and Birn$^{12}$, Wiegelmann and Schindler$^{13}$ and
Schindler$^{1}$. \\
Besides, both two- and three-dimensional ideal MHD-simulations show that the
formation of thin current sheets in the inner region of the configuration has
to be interpreted as a typical consequence of boundary
perturbations$^{14,15}$.
 \\
Further numerical simulations study the formation of current sheets in ideal
MHD configurations due to shearing of footpoints of magnetic field lines.
Strauss and Otani$^{16}$ examine the m=1 kink-ballooning mode in a magnetic
field anchored at two conducting end plates. The field lines are twisted by
shearing the footpoints at one of the boundary plates. Following the
non-linear regime of the kink instability, Strauss and Otani found the development of
current sheets with a thickness limited by resistive diffusion.
\\
Mikic \textit{et al.}$^{17}$ studied the filamentation of the current density cascading
down to smaller and smaller scales as a response to footpoint displacements in
an initially uniform magnetic field. In their simulations, however, the
plasma density is fixed and plasma pressure and resitivity are neglected.
Thus, reconnection could be possible due to uncontrolled numerical diffusion,
only.
\\
In our paper we present results of 2-D MHD-simulations 
of current filamentation due to a continued external shear
including a current density dependent resistivity, explicitly.
These
simulations allow to study both the formation and non-linear evolution of current
sheets and magnetic reconnection and its consequences. Thus, starting with an
ideal permanently externally driven plasma, we can investigate the
development of a dissipative channel and the dissipation, as well.
\\
In this context, a current density dependent resistivity is the most
realistic MHD-approach to simulate local macroscopic deviations from ideal
conductivity due to current-driven micro-instabilities. Besides, by
comparison of simulations using homogeneous and several parameter dependent
resistivity models, Ugai$^{18}$ has shown that the localization of the
resistivity is essential for a fast reconnection process.
\\
The applications we have in mind for our simulations are rather
general, namely any shear flow in collisionless astrophysical plasmas$^{19}$.
Explicitely we consider the case for extragalactic jets
stemming from active galactic nuclei. The magnetic field of extragalactic jets
is continuously sheared due to the differential rotation of the accretion disk
surrounding the central black hole. The necessary rotational energy for the
shearing is supplied by the accretion of mass onto the black hole or by
extracting the energy from a rotating black hole via the so-called
Blandford-Znajek mechanism$^{20}$. By comparing typical rotational energies
contained in the rotation of the accretion disk and the black hole$^{21}$ with
the total kinetic power of the jet$^{22}$ one can infer that no significant
slow-down of the disk's or the black hole's rotation is to be expected over a
period of the order of some Gigayears.
Extragalactic jets consist of highly
ideal low density plasmas. Observations show extended non-thermal optical and
radio emissions which require relativistic electrons accelerated up to
energies of TeV. Since the synchrotron loss lengths are considerably shorter
than the jet length the particles need to be continuously re-accelerated along
the jet$^{23,24}$.
\\
From systematical studies of the spectral indices of extragalactic jets,
Meisenheimer \textit{et al.}$^{24}$ inferred that there must be an additional
acceleration process besides local shock acceleration in the well defined
knots and hot spots. This additional acceleration process is discussed to be
magnetic reconnection$^{25,26,27}$.
\\
As the plasma is highly ideal reconnection either needs anomalous resistivity
or has to be inertia driven. Thus, current filamentation and the formation of
thin current sheets are of crucial importance with respect to jets.

\bigskip
\noindent {\bf II. The Numerical Model}

\smallskip

In the present paper, we study the process of current filamentation in sheared
magnetic fields by the help of 2-D
MHD-simulations assuming invariance in the $z$-direction ($\partial / \partial z
= 0$). \\
We use a start configuration given by a homogeneous magnetic field and a
homogeneous plasma. The external shear is realized by a rotating velocity
field~\( v \argu{r} \). Considering extragalactic jets we restrict our
2-D simulations to a plane cross section perpendicular to the jet
axis assuming an initial topology of the corresponding  field components as
simple as possible.\\
The linear and non-linear temporal evolution of the system is
then calculated by numerically integrating the following set of normalized MHD
equations
\begin{equation} \label{eq:continuity}
\dif{\rho}{t} + \nab \cdot
\argu{ \rho \vec{v} } = 0
\end{equation}
\begin{equation} \label{eq:momentum}
\dif{}{t} \argu{ \rho \vec{v} } = - \nab \cdot \argu{ \rho \vec{v} \vec{v} +
\frac{1}{2} \argu{ p + \left. \vec{B} \right. ^{2} } I_{3} - \vec{B} \vec{B} }
\end{equation}
\begin{equation} \label{eq:induction}
\dif{\vec{B}}{t} = \nab \times \argu{ \vec{v} \times \vec{B} - \eta \vec{j} }
\end{equation}
\begin{equation} \label{eq:energy}
\dif{u}{t} = - \nab \cdot \argu{ u \vec{v} } + \frac{ \gamma - 1}{ \gamma} u^{ 1
- \gamma } \eta \left. \vec{j} \right. ^{2}
\end{equation}
\begin{equation} \label{eq:ampere}
\nab \times \vec{B} = \vec{j}
\end{equation}
together with Ohm's law
\begin{equation} \label{eq:ohm}
\vec{E} + \vec{v} \times \vec{B} = \eta \vec{j}.
\end{equation}
\\
The quantity~\( u \) in the energy Equation~\eqnref{eq:energy} is given by the
relation~\( u = \argu{ p / 2 } ^{ 1 / \gamma } \). It is a measure
for the inner energy of the system. \\
The quantities~\( \rho \), \( p \), \( \vec{v} \), \( \vec{B} \), \( \vec{E} \),
\( \vec{j} \), \( \eta \), and \( \gamma \) denote the plasma mass density,
pressure, velocity, magnetic field, electric field, current density,
resistivity, and the ratio of specific heats, respectively. \( I_{3} \) denotes
the unit dyadic and \( \vec{v} \vec{v} \) represents the dyadic product \(
\argu{ v_{i} v_{j} }_{i,j} \). The ratio of specific heats has been chosen as \(
\gamma = 5/3 \). \\
All quantities~\( X \) are normalized to typical values~\( X_{0} \) of the
system. Length scales are normalized to a typical radius of a jet of~\( L_{0} = 1
\Dim{kpc} \approx 3 \cdot 10^{21} \Dim{cm} \). The particle density is
normalized to a value of~\( n_0 = 2 \cdot 10^{5} \Dim{cm^{-3}} \) while the
magnetic field is normalized to~\( B_0 = 10 \Dim{G} \) such that the values
chosen in the simulations correspond to typical core values for Active Galactic
Nuclei~(AGNs). \\
With this choice for $L_0, n_0$ and $B_0$ further normalizations follow in
a generic way, i.e. the mass density~\( \rho_{0} \), the Alfv\'en velocity~\( v
\ind{A} = B_{0} / \sqrt{4 \pi \rho_{0} } \), the Alfv\'en transit time~\( \tau
\ind{A} = L_{0} / v \ind{A} \), the electric field~\( E_{0} = v \ind{A} B_{0}
/ c \), and the resistivity~\( \eta_{0} = 4 \pi L_{0} v \ind{A} / c^{2} \)
come out to be~\( \rho_{0} \approx 1.7 \cdot 10^{-19} \Dim{g} \Dim{cm^{-3}} \)
for a quasineutral electron-proton plasma, \( v \ind{A} \approx 6.9 \cdot
10^{9} \Dim{cm} \Dim{sec^{-1}} \ll c \), \( \tau \ind{A} \approx 4.4 \cdot
10^{11} \Dim{sec} \), \( E_{0} \approx 6.9 \cdot 10^{4} \Dim{V} \Dim{m^{-1}}
\)(SI), and \( \eta_{0} \approx 2.9 \cdot 10^{11} \Dim{sec} \). \\
The
integration of the MHD equations is done on an equidistant 2-D
grid by a second order leapfrog scheme where the partial
derivatives are realized as finite differences by the FTSC method (Forward
Time Centered Space). Details of the numerical code can be found in
Otto\(^{28}\) and Otto \textit{et al.}\(^{29}\). \\ 
At the boundaries of the simulation box all
quantities with the exception of the velocity are
extrapolated to the first order of the Taylor expansion. Thus, we assume
an open plasma system where magnetic flux, plasma, and energy can freely cross
the boundaries corresponding to the fact that no generic symmetries can be defined
at the boundaries of a 2-D
cross-section of a jet.
The velocity however is assumed as a permanently given
perturbation inside the
2-D integration box.  \\
In our simulations we use a current density dependent
resistivity of the form
\begin{equation} \label{eq:resistiv2} \eta \argu{j} = \eta_{1} + \eta_{2}
\left[ j^{2} - j \ind{crit} ^{2} \right] ^{\chi} \Theta \argu{ j^{2} - j
\ind{crit} ^{2} } \end{equation}  where a small constant resistivity~\(
\eta_{1} = 10^{-5} \) has been added for numerical reasons. The coefficient~\( \eta_{2} \) is chosen to
be \( 10^{-3} \) while the critical current density~\( j \ind{crit} \) is set
to \( 1.0 \). For comparison, we also performed ideal simulations using a
small background resistivity of $10^{-5}$, only.\\
The calculations are done in a
2-D box with \( x \) and \( y \) going from \( -8 \) to \( 8 \). We
use a uniform grid with \( 461 \) grid points in the \(x\)- and in the \(y
\)-direction resulting in a constant grid spacing of about~\( 0.035 \). \\
The initial configuration is given by a homogeneous magnetic field~\( B_{0x} =
B_{0y} = 0.2 \) with no \(z\)-component (\( B_{0z} = 0 \)), a homogeneous
density~\( \rho_{0} \equiv 1.0 \), and a homogeneous~\( u_{0} \) which is given
by \( \argu{ \rho_0 / 2 }^{ 1/ \gamma } \). The initial
temperature~\( T_0 \) is homogeneous as well because of  \( u = \left[
\frac{1}{2} \rho T \right] ^{ 1 / \gamma } \) . \\
This configuration is permanently
disturbed by an externally driven rotational velocity profile \mbox{\( v_{x}
= - \sin{\phi} \; v \ind{\phi} \argu{r} \)}, \( v_{y} = \cos{\phi} \;
v \ind{\phi} \argu{r} \), with \( \phi =
\arctan{y/x} \) and~\( r = \sqrt{x^{2} + y^{2} } \). 
The azimuthal velocity is given by
\begin{equation} \label{eq:vphi}
v \ind{\phi} \argu{r} = \left\{ \begin{array}{ccc} \omega r & \mbox{for} & r \leq r
\ind{core} \\ S \ind{\Delta} \argu{r} & \mbox{for} & r \ind{core} < r < r_{1} \\
v \ind{core} \exp{ \argu{- \frac{r - r_{1}}{\lambda}}} & \mbox{for} & r_{1} \leq r
\end{array} \right. \; .
\end{equation}
The velocity profile is shown in Figure~\ref{fig:vradius0}. It corresponds to a
differential rotation with a rigid core as it is typical for accretion disks\(^{30}\).
Whereas accretion disks typically rotate with a Keplerian velocity profile~\( v
\ind{\phi} \argu{r} \sim r^{-1/2} \) outside the rigid core, in the simulations
we assume an exponentially decreasing velocity profile in order
to minimize the plasma velocity on the boundaries for numerical reasons. Using
a Keplerian velocity profile does not change the global dynamics
qualitatively.  The radius of the rigidly rotating core is given by \( r
\ind{core} = 0.3 \). The angular velocity of the rigid rotation is given
by \( \omega = v \ind{core} / r \ind{core} \) and the azimuthal velocity~\(
v \ind{core} \) at \( r \ind{core} \) is chosen to be \( 0.04 \) resulting in
a rotation period for the rigid core of about \( 47 \) Alfv\'en times. \\ The
scale length for the exponentially decreasing wing outside the radius~\( r_{1}
= 0.4 \) is given by \( \lambda = 1.5 \). Between the two radii~\( r
\ind{core} \) and \( r_{1} \) the two wings of the rotation profile are
matched by the cubic spline \begin{eqnarray}
S \ind{\Delta} \argu{r} & = & v \ind{core} + \omega \argu{r - r \ind{core}} +
\frac{1}{r_{1} - r \ind{core} } \argu{ \frac{ v \ind{core}}{ \lambda} - 2 \omega
} \argu{ r - r \ind{core} }^{2} \nonumber \\
 & & + \frac{1}{\argu{ r_{1} - r \ind{core} }^{2} }
\argu{ \omega - \frac{ v \ind{core}}{ \lambda} } \argu{ r - r \ind{core}}^{3} \; .
\end{eqnarray}
The importance of the spline is quite small as can be seen from
Figure~\ref{fig:vradius0}. It just avoids a discontinuity in the first
derivative of \( v \ind{\phi} \) regarding the radius~\( r \). \\
From systematic simulational studies we found that the dynamics of the system
is widely independent of the exact steepness of the differentially rotating
profile wing. It mainly depends on the power index in the resistivity
model~\eqnref{eq:resistiv2}. \\
In our simulations we neglect the backreaction of the jet on the accretion
disk. Therefore we restored the velocity profile after each integration step
instead of self-consistently solving the momentum equation. By this means, we
force the system to react on an unsaturated external force, represented here by
the rotation of the accretion disk.
Thus, the rotation of
the accretion disk is not slowed down during the simulation which
seems reasonable from comparing the total simulation time which is of the order
of some \( 10 \) million years with the typical age of AGN accretion disks
which is of the order of some Gigayears. \\

\bigskip
\noindent
{\bf III. Shearing of an Ideal Configuration}
\smallskip

The first example is a simulation of a configuration with external shear as
mentioned above with the plasma assumed to be ideal. There is a small
background resistivity of $\eta = 10^{-5}$ for numerical reasons, only.

\noindent
Figure~\ref{fig:maglines49} shows the structure of the initially homogeneous
magnetic field after \( \approx 1760 \; \tau \ind{A} \) which corresponds to
about \( 37 \) shear times. As to be expected, the field lines are twisted by
the differential rotation because they are frozen-in and have to
move with the plasma. The small background resistivity
allows for a very slow diffusion, only. By twisting  the
field lines several regions with antiparallel magnetic fields are created.
Going radially outward from the centre of the rotation an imaginary observer
would see a rapid change of sheets with changing direction of the magnetic
field. A sample for this observation is given in
Figure~\ref{fig:cavmagfield49} where the magnetic field vectors are presented
for the central region of the simulation box. Due to Amp\`ere's law~\( \nab
\times \vec{B} = \vec{j} \) antiparallel magnetic field lines are connected
with a current sheet. Those current sheets still have a finite but very small
thickness due to the slow diffusion. The crucial point, however, is that one
finds no evidence for some significant dissipation, especially one doesn't find any
signatures of magnetic reconnection. Up to the time scales we followed the
simulation, the system behaves as an ideal plasma.

\bigskip
\noindent
{\bf IV. Consequences of a Current Density Dependent Resistivity}

\smallskip
Considering the same
simulation with a current dependent resistivity model~\eqnref{eq:resistiv2}
(with \( \chi = 1/2 \)) the dynamics of the system changes
dramatically. \\
At the beginning phase of the simulation the system is ideal
and the field is frozen in. But
with ongoing twisting of the magnetic field, current sheets form and  the
current density in the sheets rises until it locally exceeds
the critical current density~\( j \ind{crit} \). In these
regions the anomalous resistivity not only increases the thickness of the
current sheets by diffusion  but also allows for the onset of magnetic
reconnection.  Antiparallel neighbouring
field lines are reconnected and closed field lines form around so-called
O-points. Figure~\ref{fig:maglines59} shows the magnetic field in the
non-ideal case at a similar time as in the ideal case
(Fig.~\ref{fig:maglines49}). The spiral structure of the ideal case is only
partly conserved, mainly in the outer parts of the simulation box. In the
inner region where the shear of the field lines by the differential rotation is
especially high the magnetic field lines form magnetic islands which in no way
resemble the former spiral structure. This fundamental change of the topology
of the system is a typical sign for magnetic reconnection according to
Vasyliunas\(^{10} \). \\
The regions where magnetic field lines are reconnected
coincide with the regions where the anomalous resistivity is especially high.
\\
Beside the onset and non-linear evolution of magnetic reconnection due to the
current density dependent anomalous resistivity, the formation and
dynamics of current filaments due to the shearing of the magnetic
field is a point of special interest.
(Fig.~\ref{fig:csjzc28}--\ref{fig:sjzc68}) show snapshots of the current density~\( j_{z} \)
up to
\( 4000 \) Alfv\'en-times corresponding to approximately \( 85 \)
rotations of the rigid core (i.e. more than 1 million integration timesteps).
The contours represent equidistantly distributed levels of the current
density.
 \\
Figure~\ref{fig:csjzc28} shows both a contour and a surface plot of the current
density~\( j_{z} \) at an early stage of the simulation. One finds two
developping spiral current sheets as a consequence of the external
shear. The current
density takes  its maximum at the edge
of the rigid core where the shear is the highest.
It covers a range of about \( -3
\) to \(3  \) exceeding the critical current density in a large part of
the current sheets. As a consequence of the corresponding anomalous
resistivity, the current sheets are broader as compared with the ideal case
due to diffusion. The twist of the current sheets yields
currents of alternating polarity if one goes radially
outward from the centre of the simulation box. At the outer parts of the
current sheets where the current density varies around the critical value
the sheets become unstable. Small-scale fluctuations of the
current density can be recognized in the plots already at this early time of
the simulation.
\\
A series of 8 contour plots of \(j_{z} \) at intervals of
about \( 500 \; \tau \ind{A} \) each is presented in the
Figures~\ref{fig:jzc51} and \ref{fig:jzc60}. First we remark that the
number of windings of the spiral increases with time as the spiral grows in
size. However, the growth of the spiral slows down as the outer arms come to
regions with smaller and smaller shear. Finally the configuration reaches a
state where the structure remains almost unchanged (last
pictures in Figure~\ref{fig:jzc60}). In a Keplerian rotating disk the spiral
would grow faster in time since the velocity field reaches farther out and
thus leads to a higher shear in the outer parts of the simulation box. Since
no gravitation was taken into account there is no accretion of the plasma
towards the center and the spiral is not gravitationally confined. This,
however, is not essential because similar simulations with a Keplerian rotation
have shown that the global dynamics of the filamentation process is widely
independent of the exact velocity profile. In the present example, the ``fine
structure'' of the current sheets, i.e.~the filaments are defined by the value
of the critical current density and the exponent~\( \chi \) in the resistivity
model~\eqnref{eq:resistiv2}. The higher the exponent~\( \chi \) the faster the
filamentation of the system. Simulations have been done with \( \chi = 1/4\),
\( 1/2 \), and \( 1 \) yielding that the time scales for the filamentation and
the tearing of the current sheets become smaller with increasing \( \chi \).
By comparing the different simulations, one finds that the maximum resistivity in
the reconnection zones increases with increasing \( \chi \).
Therefore, the behaviour of the time scale for the filamentation is in
qualitative agreement with the theory where the typical time
scale for the magnetic reconnection \( \tau \ind{A} N \ind{L} ^{1/2} \)
decreases with increasing resistivity~\( \eta \) and thus decreasing Lundquist
number~\( N \ind{L} = 4 \pi L v \ind{A} / c^{2} \eta \).
\\
The
sequence of contour plots shows that the spiral current sheets become rapidly
unstable. Starting at the borders of the current sheet numerous small
scale fluctuations around \( j_{z} = 0 \) develop. These small scale current
filaments are a consequence  of the localization of the anomalous
resistivity.
\\
Besides, the
sequence  shows that the current system is unstable with respect to
reconnection  and the coalescence instability. At \( \approx 2000
\tau \ind{A} \) \bref{fig:jzc51}{d}, for instance, one inner winding of the
spiral current sheet has been torn apart due to reconnection.
The disrupted current sheet however becomes unstable against the
coalescence instability, and  at about \( 3000 \tau \ind{A} \)
\bref{fig:jzc60}{b} the broken winding is re-merged.
\\
Similarly, the small
scale filaments are unstable to the coalescence instability due to the
attraction of  parallel currents. Via the coalescence of
several small scale filaments large scale filaments are created. Some samples
of large scale current filaments can be seen on the right- and left-hand side
of the spiral from about \( 1500 \tau \ind{A} \) onwards. At later times (\(>
3000 \tau \ind{A} \)) similar current filaments show up at the top and the
buttom of the spiral. Those large scale filaments grow in length thinning
more and more until they become unstable against reconnection.
The maximum extension in length is about 5 -- 6 unit lengths.
In the course of time the current filaments undergo alternating phases
of reconnection and coalescence both
converting magnetic energy into heat and kinetic energy of particles
due to  Ohmic dissipation and acceleration via electric fields~\(
E_{z} \), respectively. The large scale current
filaments more or less keep their spatial position even while the spiral keeps
growing. Since the pressure gradient~\( \nab p \) is negligible the filaments
can be considered to be force-free. Concerning the energy transport in a jet
such force-free filaments could provide a powerful mechanism of
particle acceleration and re-acceleration via magnetic reconnection.
\\
Figure~\ref{fig:sjzc68}  shows a
surface plot of the current density~\( j_{z} \) for the
last timestep of the previous sequence, i.e. $t = 4000 \tau_A$. One finds
that the spiral structure is almost completely disrupted by current filaments,
some of them even separated from the former current sheets.
\\
The filamentary structure of the current system can also be
found in a contour plot of the current density dependent resistivity
\aref{fig:resc68}. The size of the
filaments approaches the smallest scales resolved in the simulation, i.e.
the grid cell. Thus, one can expect that the process of filamentation would
continue to even smaller scales if the spatial resolution was increased. Such
a cascade to smaller and smaller scales may be important with respect towards
inertia driven magnetic reconnection\(^{5,10}\).
The resistivity ranges between the constant background value of \( 10^{-5}
\) and the peak value of about \( 2.5 \cdot 10^{-3} \). The current filaments
with the highest resistivity simultaneously represent the regions with the
highest Ohmic dissipation~\( \eta \left. \vec{j} \right. ^{2} \) and the
largest electric field~\( E_{z} \). Thus they are the most promising
candidates for plasma heating and particle acceleration.
\\
A further point of interest considering the dynamics of the system is
the question on the temporal evolution of the magnetic
energy. For these purposes we integrate the magnetic energy
density~\( 1/2\left. \vec{B} \right. ^{2} \) over the whole simulation box for
each time step and plot the result versus the time \aref{fig:magenergy}. The
initial magnetic energy density corresponds to the homogeneous magnetic field
of the start configuration. By conversion of kinetic shear flow energy
into magnetic energy via the generator term \( \vec{v} \cdot \argu{ \vec{j}
\times \vec{B} } \) in the normalized equation for the magnetic energy$^{31}$ 
\begin{equation}  \label{eq:magnenergy}
- \frac{1}{2} \dif{}{t} \left. \vec{B} \right. ^{2} = \eta \left. \vec{j}
  \right. ^{2} + \vec{S} + \vec{v} \cdot \argu{ \vec{j} \times \vec{B} }
\end{equation}
the magnetic energy of the system rapidly increases for about \(
1000 \) Alfv\'en times. Here \( \vec{S} = \vec{E} \times \vec{B} \) denotes the
Poynting flux. Then, the formation of current sheets and
the occurence of an anomalous resistivity lead to
dissipation due to reconnection. Magnetic  energy
is converted into heat via Ohmic dissipation and into kinetic energy via
particle acceleration. The plasma flow acceleration by magnetic reconnection
however is not resolved quantitatively because of the external driving. Still,
the resulting electric fields~\( E \ind{z} \) allow for the acceleration of test
particles along the separator lines. Because of the dissipation the slope of the 
magnetic energy~\( E \ind{mag} \argu{t} \) decreases. Finally the system
reaches a kind of energy balance after \( \approx 2500 \tau \ind{A} \). In this
state energy is continuously transferred via filamentation from large scales to
small scales where it is dissipated. The dissipation, however, leads to a
continuous increase of the thermal energy~\( U=3p/2 \) of the system
(Figure~\ref{fig:thermal}) because radiative losses
are not included in the energy Equation~\eqnref{eq:energy}.\\
Another interesting point is the temporal evolution of the
total Ohmic dissipation rate of the system \( P \ind{diss} \argu{t} =
\int_{\mathcal{F}}^{} \eta \left. \vec{j} ^{2} \right. dxdy \), where \(
\mathcal{F} \) denotes the simulation box. Figure~\ref{fig:dissrate} shows the
dissipation rate~\( P \ind{diss} \) versus the time~\( t \). At the very
beginning of the simulation the current density is zero, i.e.~\( P \ind{diss}
\argu{0} = 0 \). For the first \( 500 \) Alfv\'en-times the dissipation rate
grows almost linearly due to the formation of large scale current sheets and
the onset of anomalous resistivity. But then the dissipation
rate stagnates and decreases a little bit until it grows again but much more
slowly. Considering the fine structure of the
dissipation rate one finds that it becomes more and more noisy
with the course of time.
This is in qualitative agreement with results from
Haswell \textit{et al.}\(^{32}\) who studied a 
similar system but with a homogeneous resistivity where the creation and
dissipation of magnetic energy in a differentially rotating non-ideal system
takes place periodically. Magnetic energy is built up by the external shear and
then spontaneously dissipated  by magnetic reconnection. In our simulations
magnetic reconnection
occurs only quasiperiodically because we consider a current density dependent
resistivity. Especially, the scales for the dissipation scale from the large
scales of the global current sheets to the small scales of single current
filaments. A 3-D-animation of the current density~\( j_{z} \) shows a pulsating
shape with small-scale eruptions like on the surface of a boiling stew. Those
temporal fluctuations of \( j_{z} \) are responsible for the noisy structure
of the curve~\( P \ind{diss} \argu{t} \). The first fluctuations until \(
\approx 1200 \tau \ind{A} \) are large-scale fluctuations of the current
density. They are replaced by the small-scale fluctuations of the current
filaments being formed in the course of the simulation. In the course of time
the curve becomes more and more noisy reflecting the progressive filamentation
of the current system. The temporal resolution of the simulation is given by
the timestep of \( 0.0042 \tau \ind{A} \) such that even finest temporal
fluctuations are resolved. Thus, the  Ohmic dissipation rate also
clearly shows the ongoing filamentation. \\

\bigskip
\noindent
{\bf V.  Discussion}

\smallskip
In our paper we discussed 2-D MHD-simulations of current
filamentation in an initially globally ideal plasma due to permanently
externally driven shear flows.
\\
The configuration under consideration can be interpreted as cross sections of
extragalactic jets perpendicular to the jet axis. Thus, the magnetic field
component along the jet axis has been neglected. The driving shear flow is the
consequence of differential rotation in the accretion disk surrounding a
central black hole.
\\
We discussed and compared two simulational studies. In the first example, we
studied the dynamics of a shear flow driven system being globally ideal for
the whole simulation time. Thus, there were no dissipative channels available
to dissipate the energy pumped into the system due to the external force. As a
consequence, one finds an ongoing twisting of the magnetic field together with
the development of current sheets.
\\
\\
In the second example we studied the same configuration as in example one, but
now assuming a current density dependent resistivity. This allows to mimic
the development of a dissipative channel due to current-driven
micro-instabilities yielding local non-idealness (anomalous resistivity) on
macroscopic scales.
\\
As soon as the current density dependent resistivity is switched on locally,
the system can dissipate the external energy effectively by magnetic
reconnection.   As a consequence, the dynamics of the system differs
significantly from that found in the globally ideal case.
\\
In the beginning one finds a twisting of the magnetic field and developping
current sheets as in example one. Later, however, the dynamics is dominated by
the tearing of current sheets due to reconnection and the merging of current
filaments due to coalescence instabilities.
\\
In the course of that, microscopic structures cascade down to smaller and
smaller scales and magnetic energy is converted into heat and kinetic
particle energy. From an energetical point of view, the system finally reaches
a state with an energy balance between external input and
internal dissipation. The resulting large scale current filaments can be
considered to be force-free.
\\
\\
Considering the observed radio emissions in extragalactic jets these current
sheets could provide a powerful source for local non-idealness and for continuous
acceleration and re-acceleration of electrons up to TeV energies due to
magnetic reconnection. Besides, the net current inside an extragalactic jet
necessary to collimate the configuration has to be of the order of typically
$10^{19}$ Amp\`ere, whereas a single current layer cannot carry currents higher than
the Alfv\'en limit of $\gamma \cdot 10^4$ Amp\`ere. Thus, the total current inside a jet
has to be carried by a system of multi-current layers which could be provided
by current filamentation as found in our simulations.
\\
\\
The simulations discussed in the present paper are 2-D ones
without taking into account the magnetic field parallel to the jet axis.
Corresponding 3-D simulations including the dynamics along the
jet axis will be the subject of future work.

\bigskip\bigskip

\begin{center} {\bf ACKNOWLEDGEMENTS}
\end{center}

\bigskip

\noindent
This work was supported by the
Deutsche Forschungsgemeinschaft through the grants LE~1039/3-1,5-1. \\
We thank the referee for his helpful comments.

\newpage

\begin{center} {\bf REFERENCES}
\end{center}
\newcounter{no}
\begin{list}
{[\arabic{no}]}{\usecounter{no}}

\item Schindler, K., Physica Scr. {\bf T50}, 20 (1993)

\item Parker, E.N., in {\it Solar and Astrophysical Magnetohydrodynamic
Flows}, ed. by K.C. Tsinganos (Kluwer, Dordrecht), 337 (1996)

\item Hayashi, M.R., K. Shibata, and R. Matsumoto, Astrophys.~J. {\bf 468},
L37 (1996)

\item Lesch, H. and Reich, W., Astron. Astrophys. {\bf 264}, 493 (1992)

\item Birk, G.T. and Lesch, H., Astrophys.~J. {\bf 530}, L77 (2000)

\item Lesch, H. and G.T. Birk, Astrophys.~J. {\bf 499}, 167 (1998)

\item Petschek, H.E., in {\it AAS-NASA Symposium on Physics of Solar Flares},
NASA Spe. Publ. {\bf 50} (National Aeronautics and Space Administration,
Washington DC, 1964), p.~425

\item Parker, E.N., in {\it Spontaneous Current Sheets in Magnetic Fields}
(Oxford University Press, New York, 1994)

\item Huba, J.D., in {\it Unstable Current Systems and Plasma Instabilities in
Astrophysics, IAU 107}, ed. by M.R. Kundu and G.D. Holmann (Reidel, Dordrecht, 
1985), p. 315

\item Vasyliunas, V.M., Theoretical Models of Magnetic Field Line Merging, 1,
Rev. of Geophys. and Space Phys. {\bf 13}, 303 (1975)

\item Hahm, T.S. and R.M. Kulsrud, Physica Scr. {\bf 2}, 525 (1982)

\item Schindler, K. and J. Birn, J. Geophys. Res. {\bf 98}, 477 (1993)

\item Wiegelmann, T. and K. Schindler, Geophys. Res. Lett. {\bf 15}, 2057
(1995)

\item Wiechen, H., Ann. Geophysicae {\bf 17}, 595 (1999)

\item Wiechen, H., G.T. Birk, and H. Lesch, Phys. Plasmas {\bf 5}, 3732 (1998)

\item Strauss, H.R. and N.F. Otani, Astrophys.~J. {\bf 326}, 418 (1988)

\item Mikic, Z., D.D. Schnack, and G. van Hoven, Astrophys.~J. {\bf 338}, 1148
(1989)

\item Ugai, M., Phys. Fluids B {\bf 4} (9), 2953 (1992)

\item Lesch, H. in \textit{Solar and Astrophysical MHD-Flows}, ed.~K.~Tsinganos,
NATO-ASI-Series {\bf 481} (Kluwer Academic Publishers, Dordrecht,1996), p.~ 673

\item Blandford, R.D. and Znajek, R.L., MNRAS {\bf 179}, 433 (1977)

\item Camenzind, M. in \textit{Theory of Accretion Disks - 2}, ed.~by
W.J.~Duschl, J.~Frank, F.~Meyer, E.~Meyer-Hofmeister, and W.M.~Tscharnuter, NATO
ASI-Series, Series C: Mathematical and Physical Sciences - {\bf Vol.~417}, 313--328

\item Celotti, A., Kuncic, Z., Rees, M.J., and Wardle, J.F.C., MNRAS {\bf 293},
288--298 (1998)

\item Begelman, M.C., R.D. Blandford, and M.J. Rees, Rev. Mod. Phys. {\bf 56},
255 (1984)

\item Meisenheimer, K., M.G. Yates, and H.-J. R\"oser, Astron. Astrophys. {\bf
325}, 57 (1997)

\item Romanova, M.M., and R.V.E. Lovelace, Astron. Astrophys. {\bf 262}, 26
(1992)

\item Vekstein, G.E., E.R. Priest, and C.D.C. Steele, Astrophys.~J. Suppl.
{\bf 92}, 111 (1994)

\item Blackman, E.G., Astrophys.~J. {\bf 456}, L87 (1996)

\item Otto, A., Comp. Phys. Com. {\bf 59}, 185 (1990)

\item Otto, A., Schindler, K., Birn, J., J. Geophys. Res. {\bf 95}, 15023 (1990)

\item Frank, J., King, A.R., Raine, D.J. in \textit{Accretion Power in
Astrophysics}, (Cambridge University Press, Cambridge, 1992)

\item Kippenhahn, R. and M\"ollenhoff, C. in \textit{Elementare Plasmaphysik},
(B.I.-Wissenschaftsverlag, Z\"urich, 1975), p.~98

\item Haswell, C.A., Tajima, T., Sakai, J.-I., Astrophys.~J. {\bf 401}, 495 (1992)

\end{list}

\newpage

\noindent {\bf FIGURE CAPTION}

\bigskip
\noindent Fig.1 Possible scenarios for the physical system under investigation.

\bigskip
\noindent Fig.2 The azimuthal velocity~\( v \ind{\Phi} \argu{r} \) for the
  2-D simulations.

\bigskip
\noindent Fig.3 The magnetic field lines in an ideal plasma sheared by a
  differential rotation at \( t = 1760.53 \; \tau \ind{A} \).

\bigskip
\noindent Fig.4 Regions with antiparallel magnetic field vectors in the ideal
  case after \( 1760.53 \; \tau \ind{A} \) shearing time.

\bigskip
\noindent Fig.5 Reconnected field lines in the non-ideal case at \( t = 1759.72 \;
  \tau \ind{A} \).

\bigskip
\noindent Fig.6 Current density~\( j_{z} \) for the non-ideal case at \( t = 100.78 \;
 \tau \ind{A} \).

\bigskip
\noindent Fig.7 Series of contour plots of the current density~\( j_{z} \) for the
 non-ideal case at different times.

\bigskip
\noindent Fig.8 Series of contour plots of the current density~\( j_{z} \) for the
 non-ideal case at different times.

\bigskip
\noindent Fig.9 Surface plot of the current density~\( j_{z} \) at the end of the
  simulation (\(4001.3 \; \tau \ind{A} \)).

\bigskip
\noindent Fig.10 Contour plot of the resistivity~\( \eta \) at the end of the
  simulation (\(4001.3 \; \tau \ind{A} \)).

\bigskip
\noindent Fig.11 Temporal evolution of the magnetic energy of the system.

\bigskip
\noindent Fig.12 Temporal evolution of the thermal energy of the system.

\bigskip
\noindent Fig.13 Temporal evolution of the Ohmic dissipation rate of the system.

\newpage

 \newlength{\picwid}
 \setlength{\picwid}{0.95\linewidth}

\begin{figure}[t]
  \center
  \includegraphics[angle=0,width=\picwid,keepaspectratio]%
	{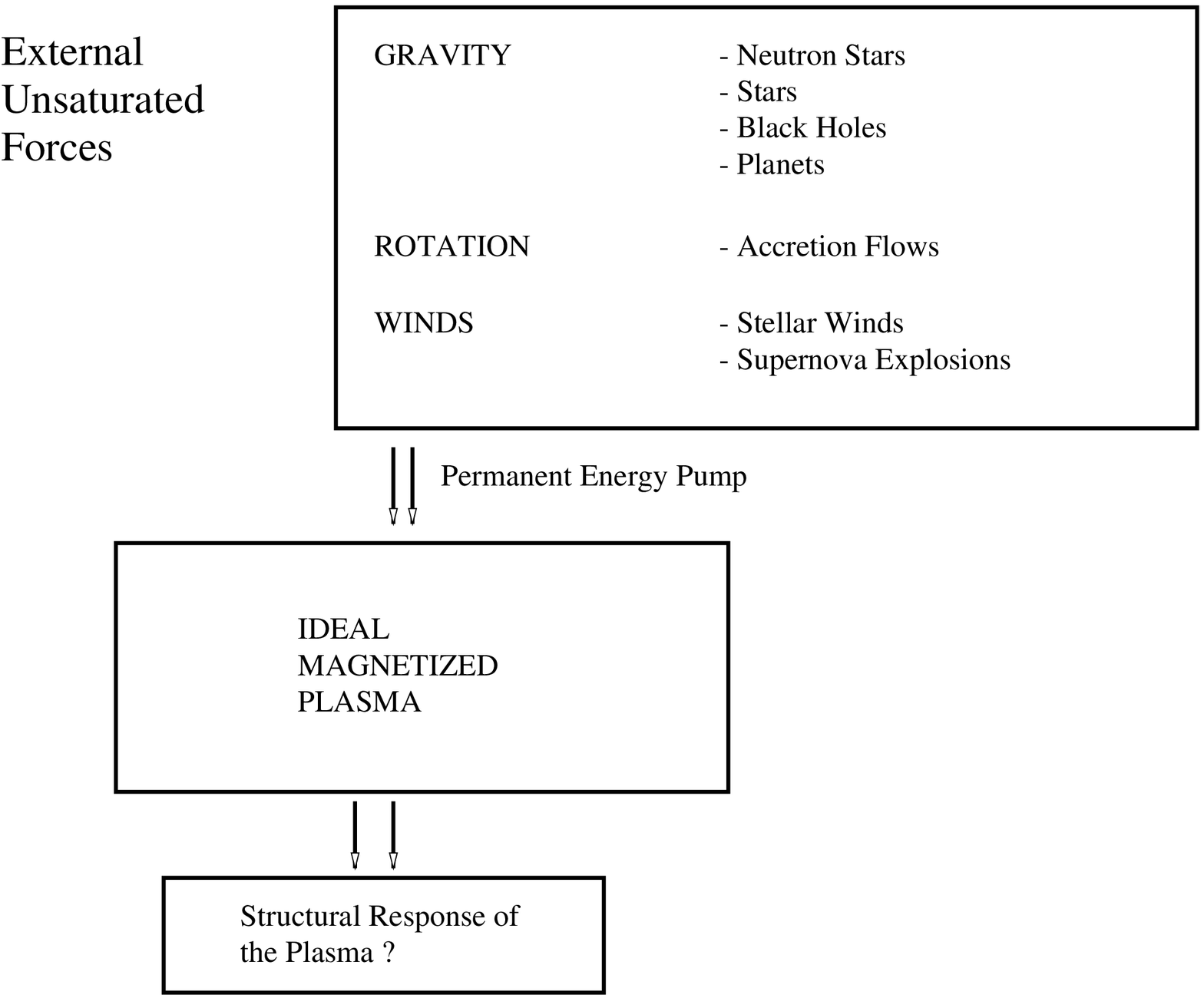}
  \caption{Possible scenarios for the physical system under investigation}
  \label{fig:intention}
\end{figure}

\newpage

\begin{figure}[t]
  \center
  \includegraphics[angle=90,width=\picwid,keepaspectratio]%
	{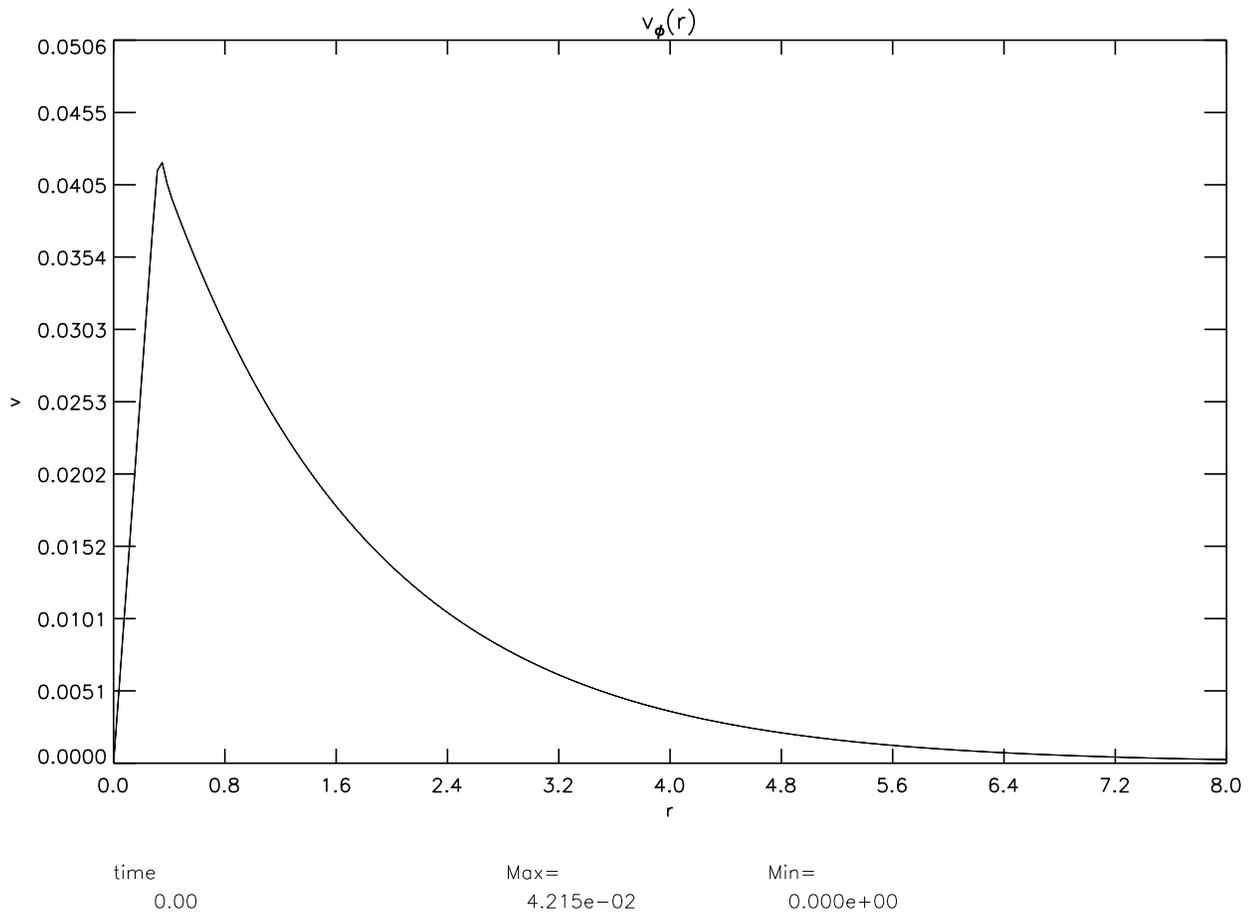}
  \caption{The azimuthal velocity~\( v \ind{\Phi} \argu{r} \) for the
  2-D simulations }
  \label{fig:vradius0}
\end{figure}

\newpage

\begin{figure}[t]
  \center
  \includegraphics[angle=90,width=\picwid,keepaspectratio]%
	{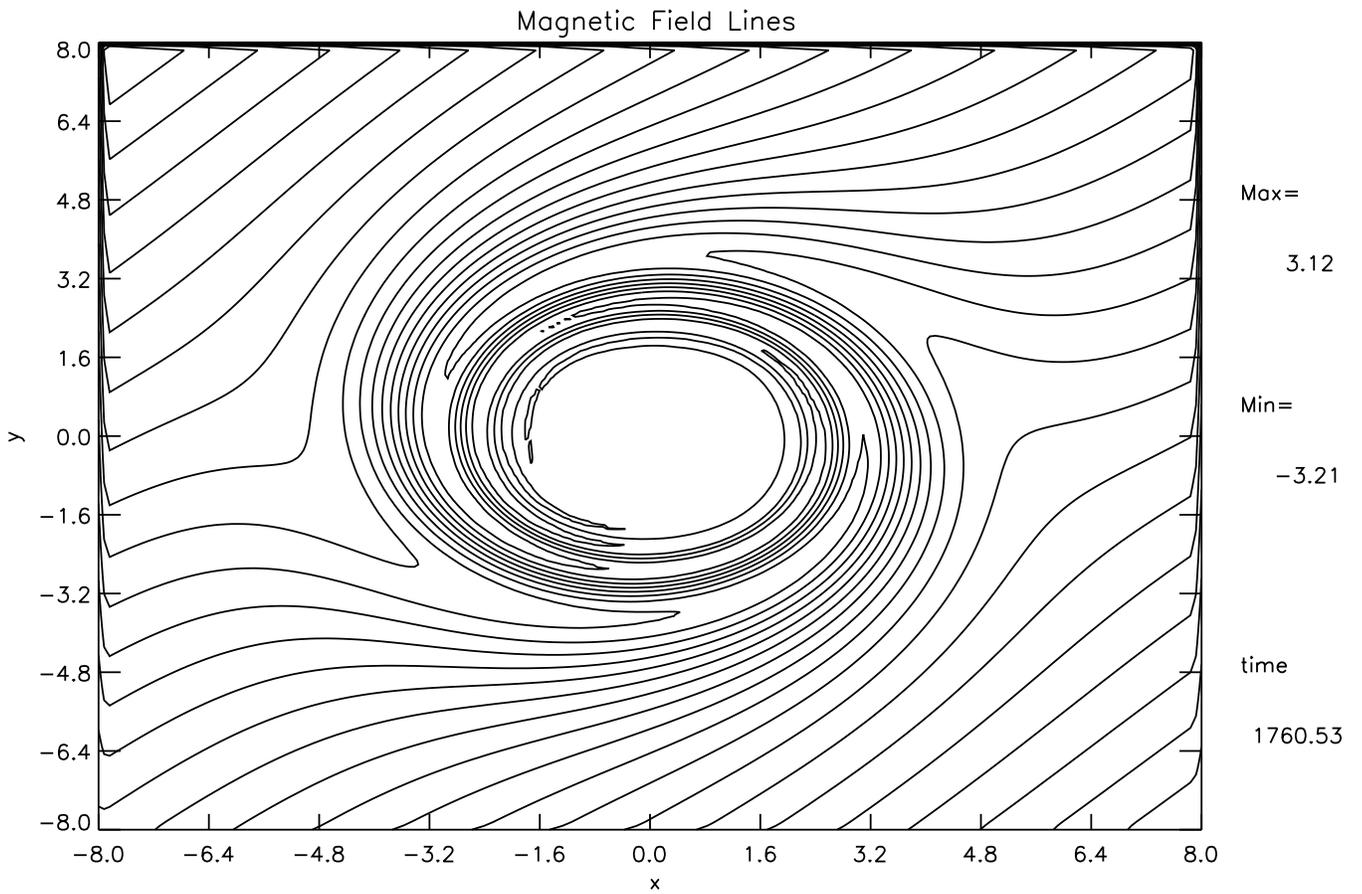}
  \caption{The magnetic field lines in an ideal plasma sheared by a differential
  rotation at \( t = 1760.53 \; \tau \ind{A} \) }
  \label{fig:maglines49}
\end{figure}

\newpage

\begin{figure}[t]
  \center
  \includegraphics[angle=90,width=\picwid,keepaspectratio]%
	{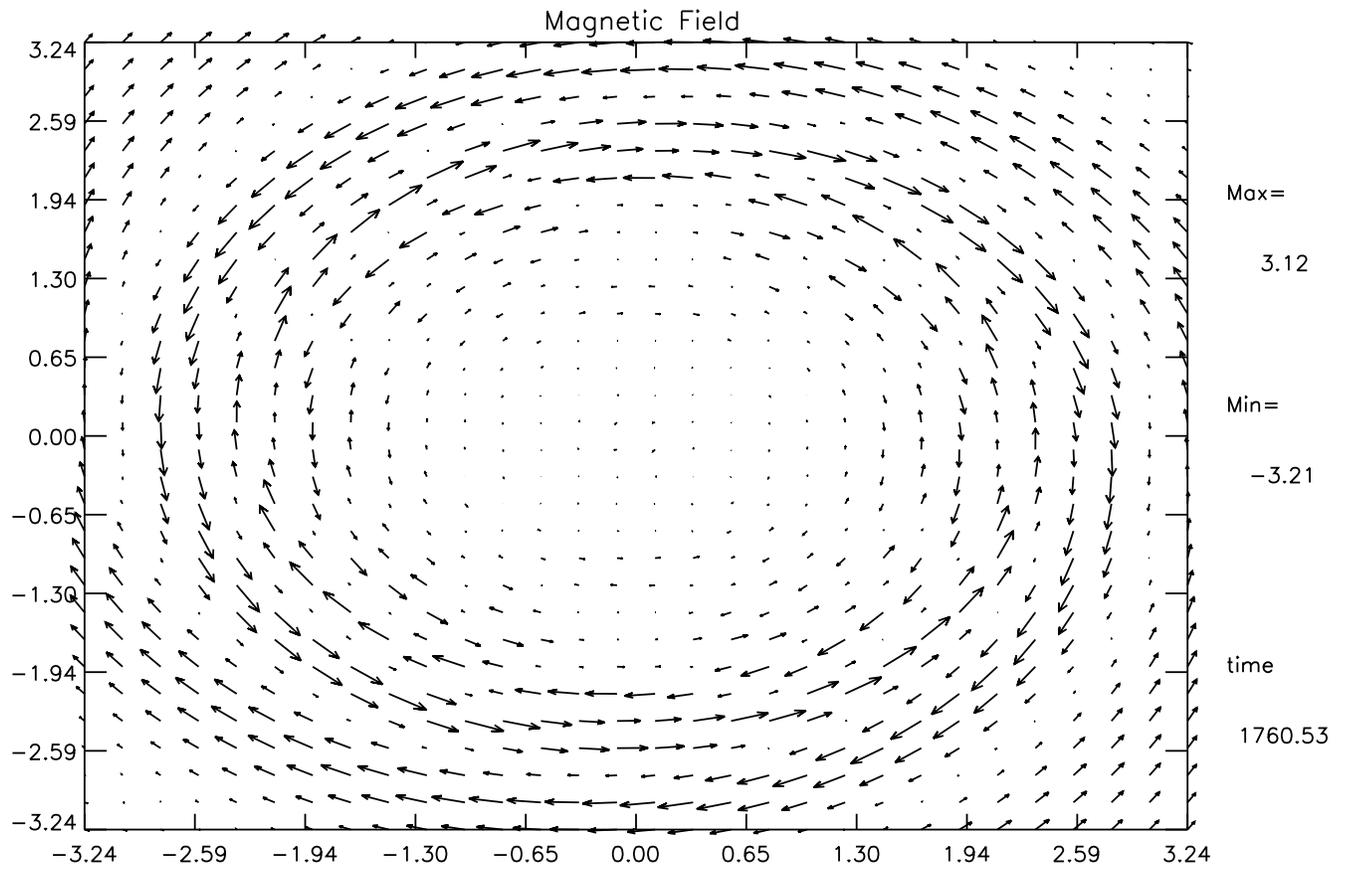}
  \caption{Regions with antiparallel magnetic field vectors in the ideal case
  after \( 1760.53 \; \tau \ind{A} \) shearing time}
  \label{fig:cavmagfield49}
\end{figure}

\newpage

\begin{figure}[t]
  \center
  \includegraphics[angle=90,width=\picwid,keepaspectratio]%
	{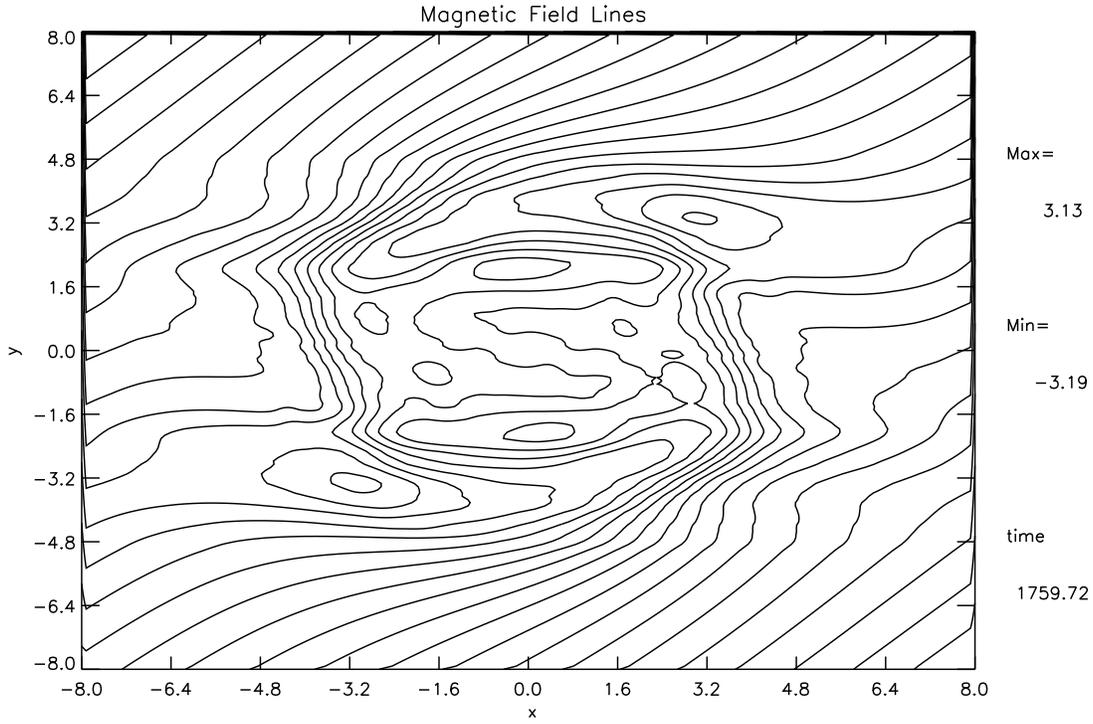}
  \caption{Reconnected field lines in the non-ideal case at \( t = 1759.72 \;
  \tau \ind{A} \)}
  \label{fig:maglines59}
\end{figure}

\newpage

\begin{figure}[t]
 \centering
 \subfigure[Contour plot]{\includegraphics[angle=90,width=0.8\picwid,
 keepaspectratio]{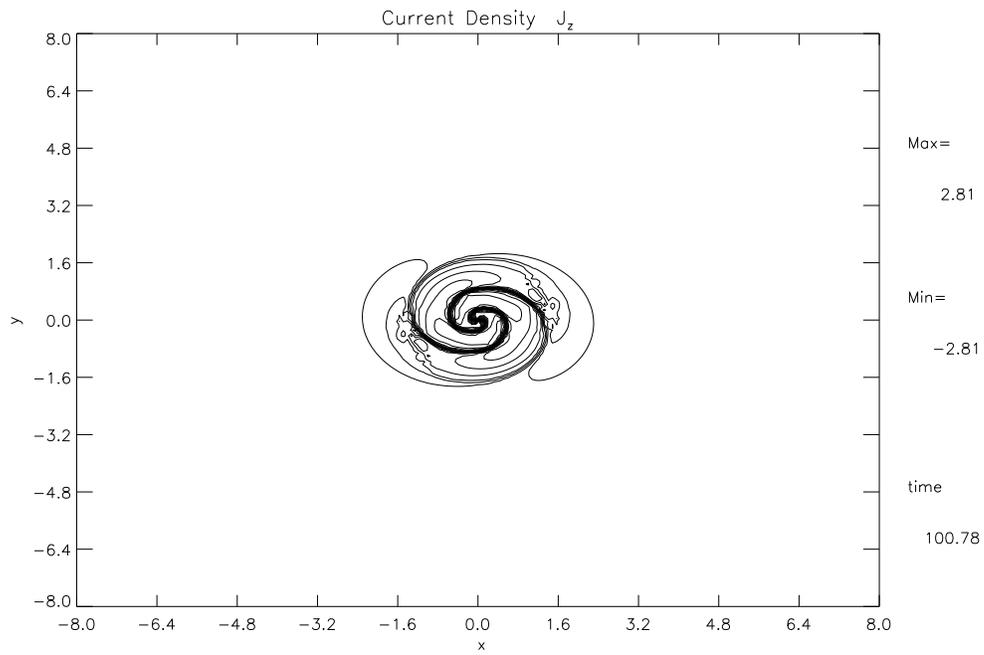}} \\
 \subfigure[Close-up surface plot]{\includegraphics[angle=90,width=0.8\picwid,
 keepaspectratio]{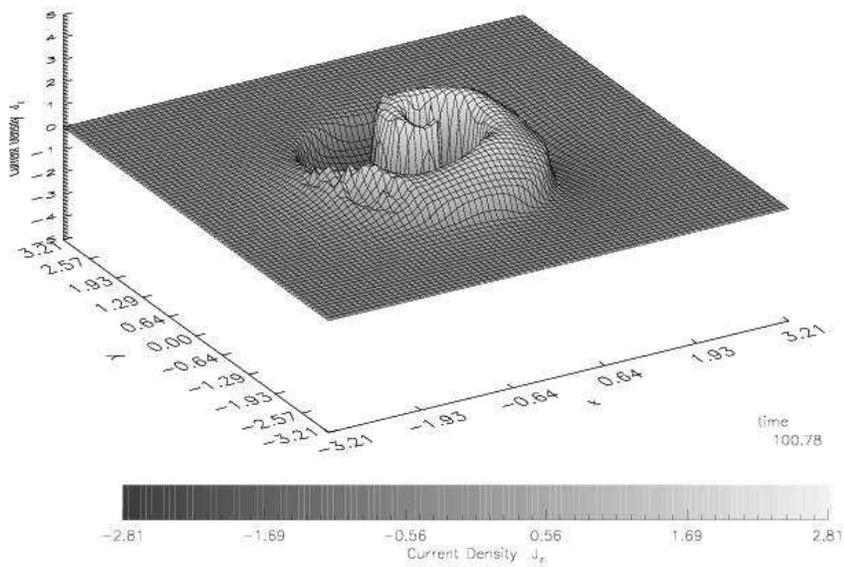}}
 \caption{Current density~\( j_{z} \) for the non-ideal case at \( t = 100.78 \;
 \tau \ind{A} \)}
 \label{fig:csjzc28}
\end{figure}

\newpage

\begin{figure}[t]
 \centering
 \subfigure[\( t = 499.32 \; \tau \ind{A} \)]{\includegraphics[angle=90,
 width=0.48\picwid, keepaspectratio]{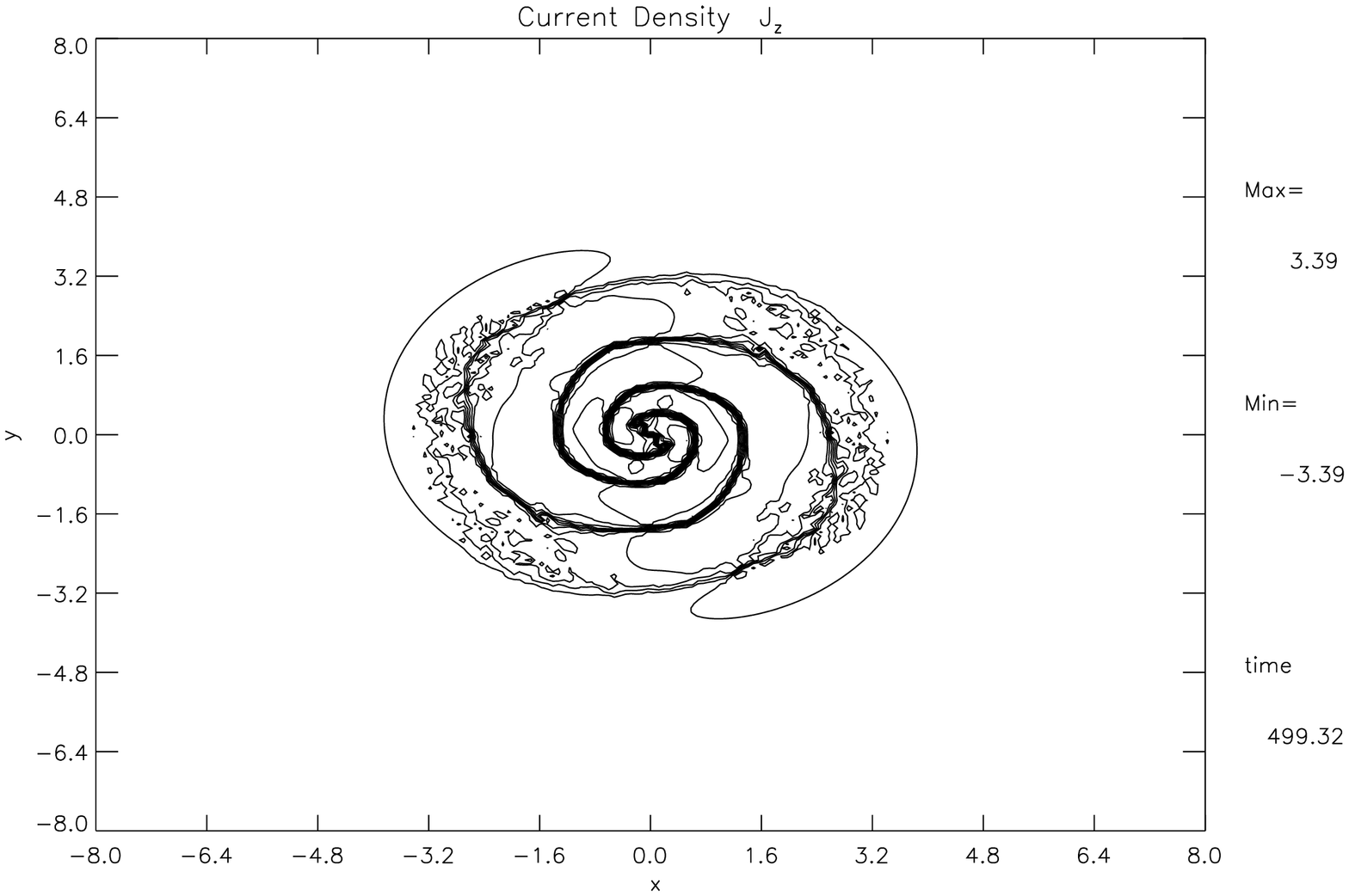}} \quad
 \subfigure[\( t = 999.72 \; \tau \ind{A} \)]{\includegraphics[angle=90,
 width=0.48\picwid, keepaspectratio]{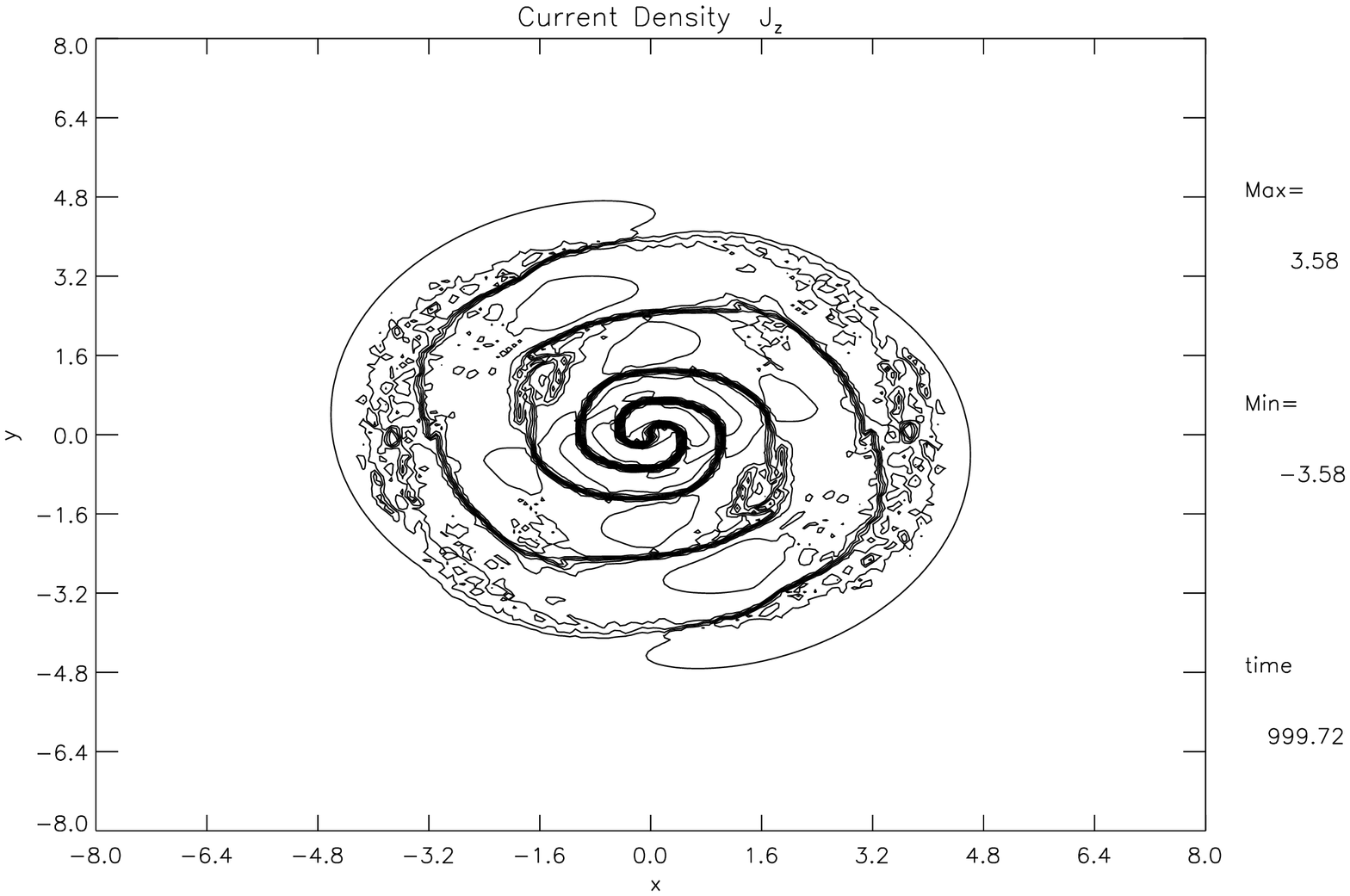}} \\
 \subfigure[\( t = 1500.69 \; \tau \ind{A} \)]{\includegraphics[angle=90,
 width=0.48\picwid, keepaspectratio]{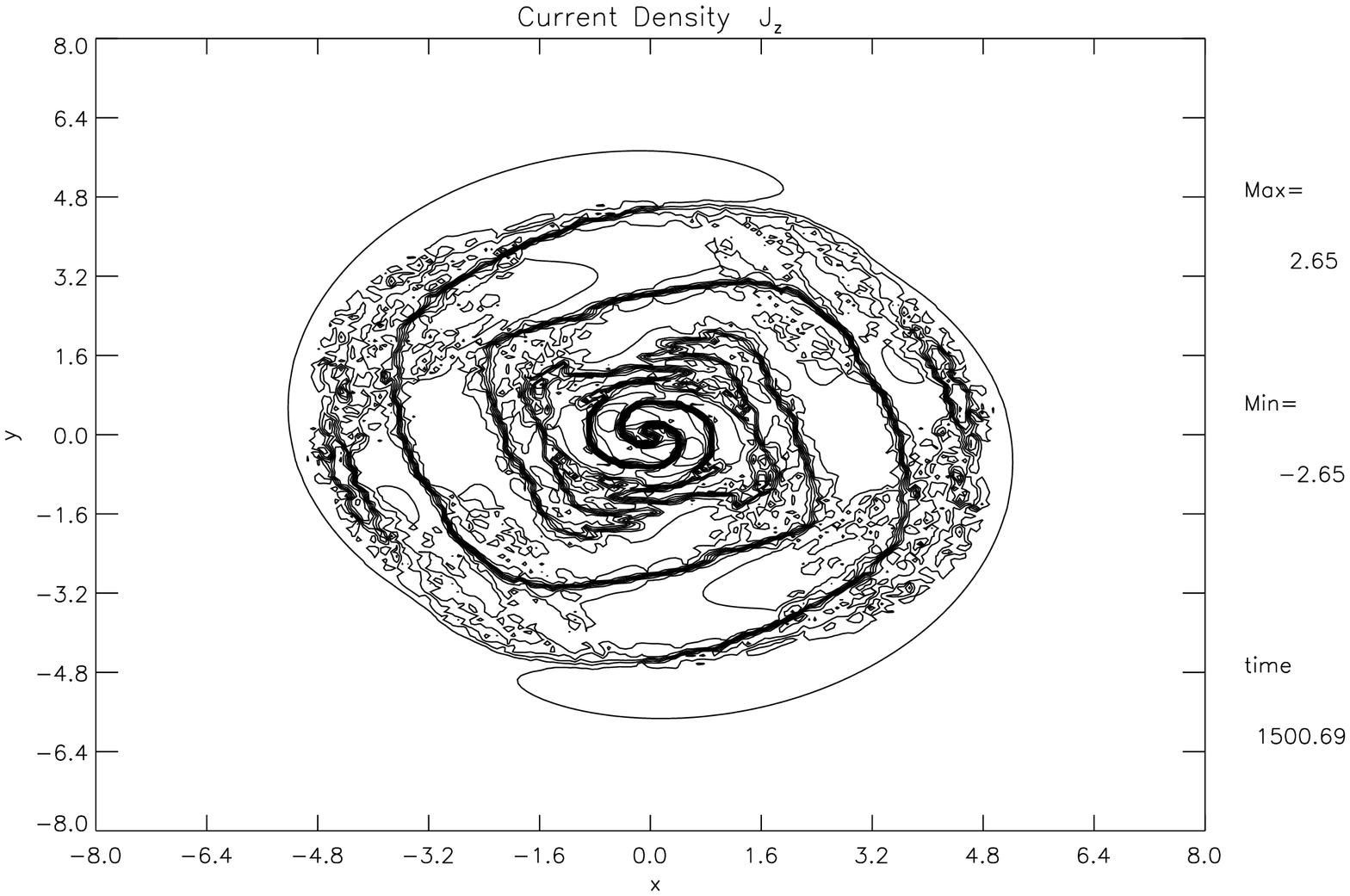}} \quad
 \subfigure[\( t = 2001.31 \; \tau \ind{A} \)]{\includegraphics[angle=90,
 width=0.48\picwid, keepaspectratio]{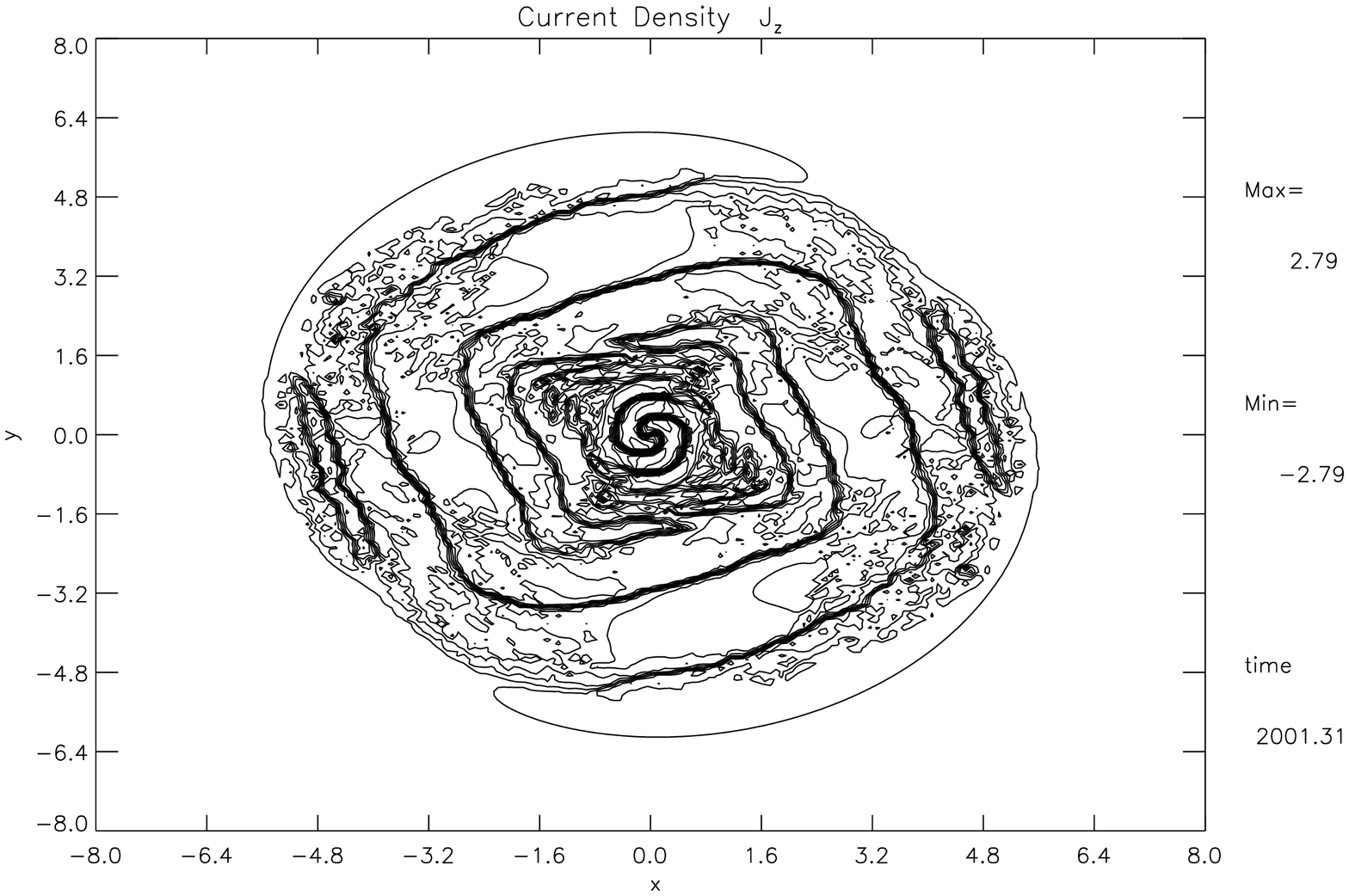}}
 \caption{Series of contour plots of the current density~\( j_{z} \) for the
 non-ideal case at different times}
 \label{fig:jzc51}
\end{figure}

\newpage

\begin{figure}[t]
 \centering
 \subfigure[\( t = 2501.93 \; \tau \ind{A} \)]{\includegraphics[angle=90,
 width=0.48\picwid, keepaspectratio]{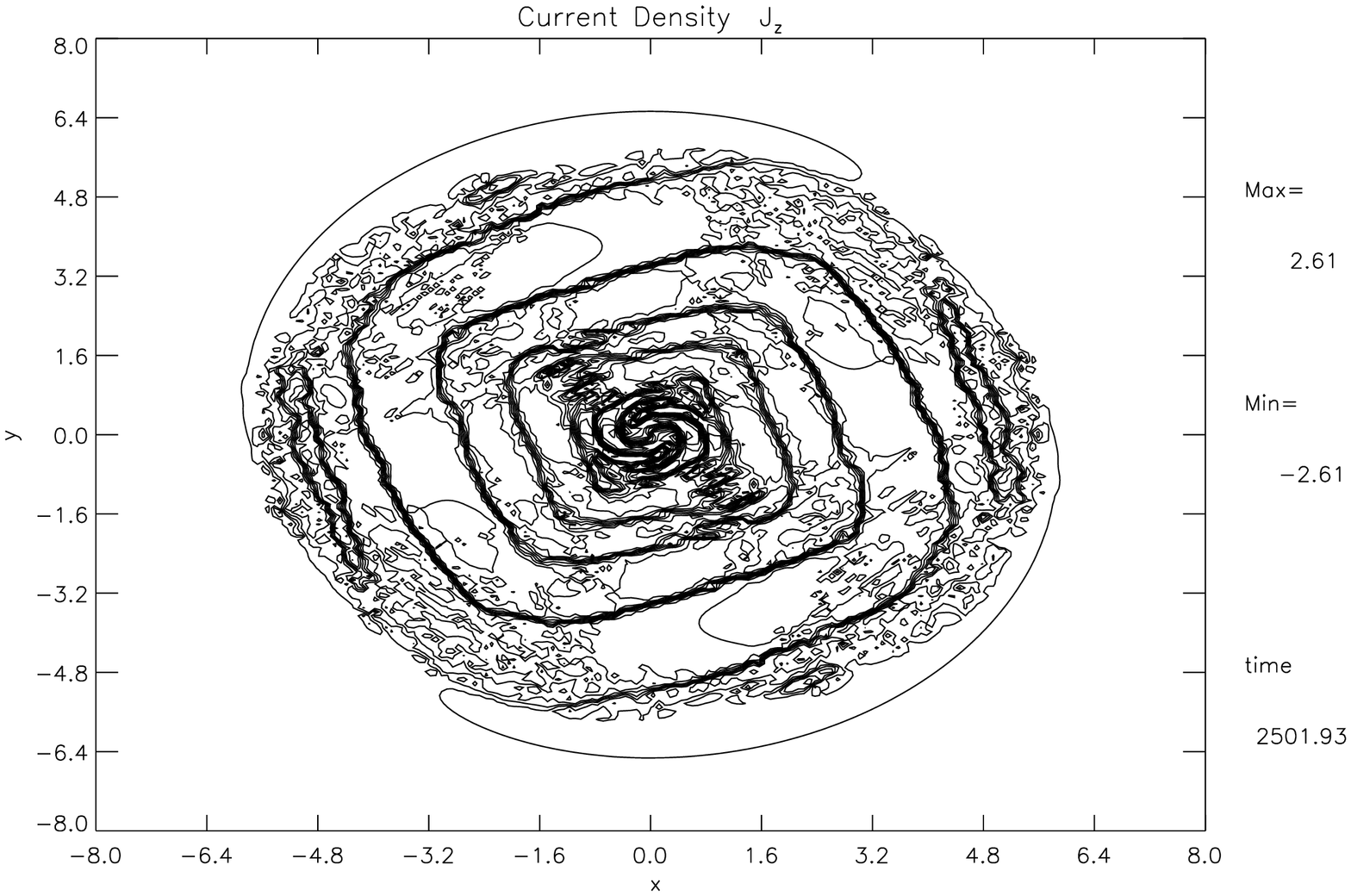}} \quad
 \subfigure[\( t = 3000.06 \; \tau \ind{A} \)]{\includegraphics[angle=90,
 width=0.48\picwid, keepaspectratio]{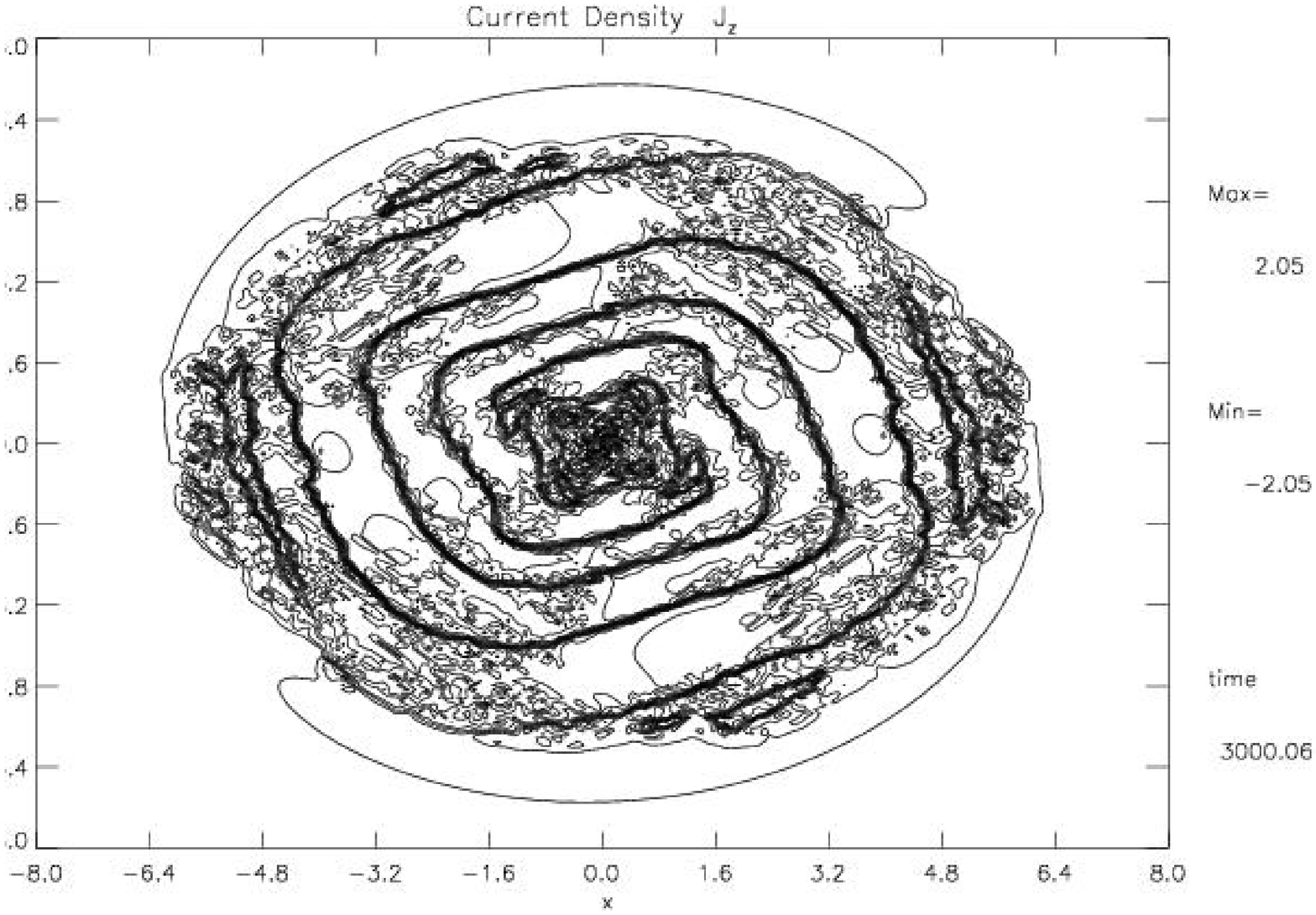}} \\
 \subfigure[\( t = 3500.68 \; \tau \ind{A} \)]{\includegraphics[angle=90,
 width=0.48\picwid, keepaspectratio]{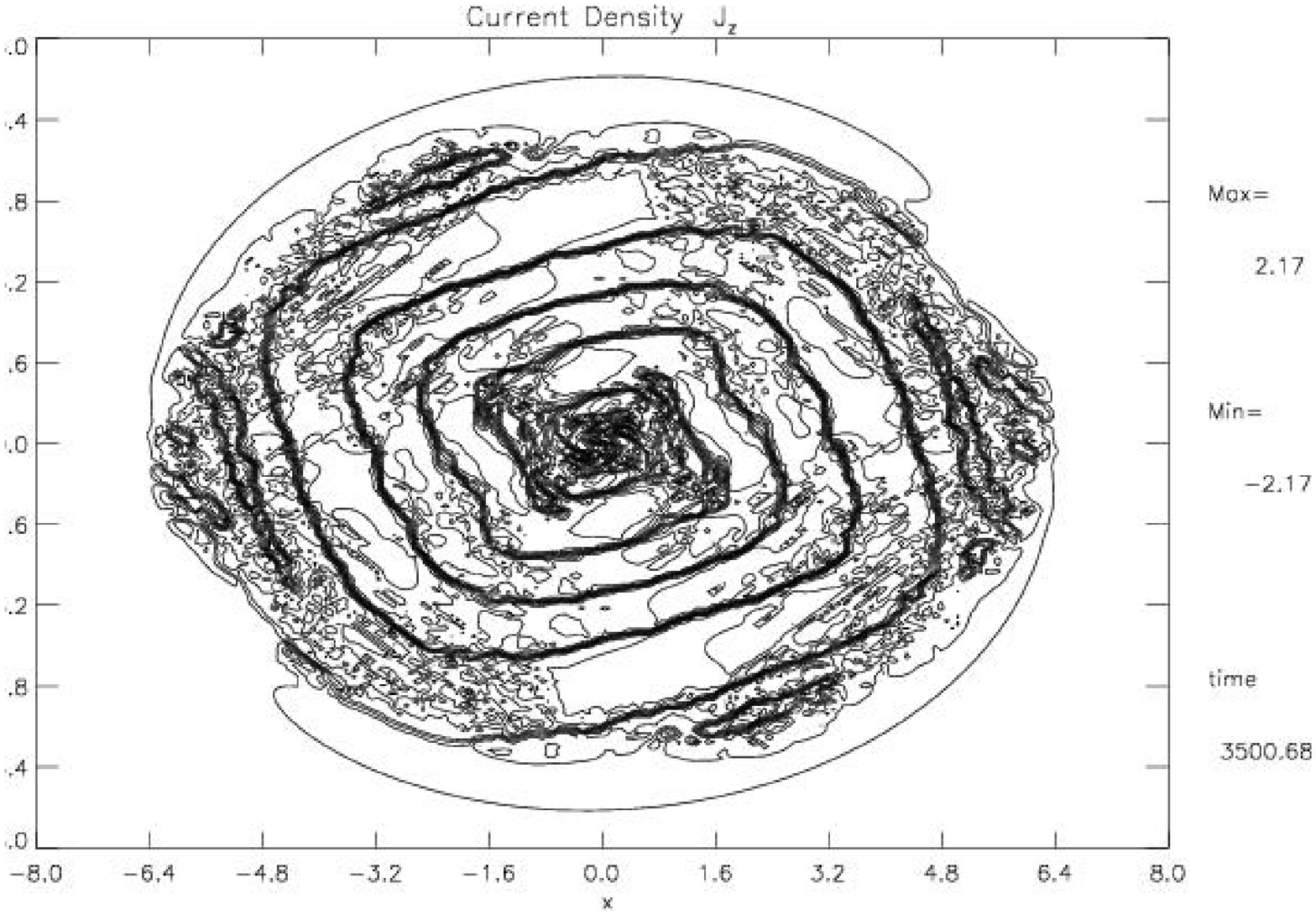}} \quad
 \subfigure[\( t = 4001.3 \; \tau \ind{A} \)]{\includegraphics[angle=90,
 width=0.48\picwid, keepaspectratio]{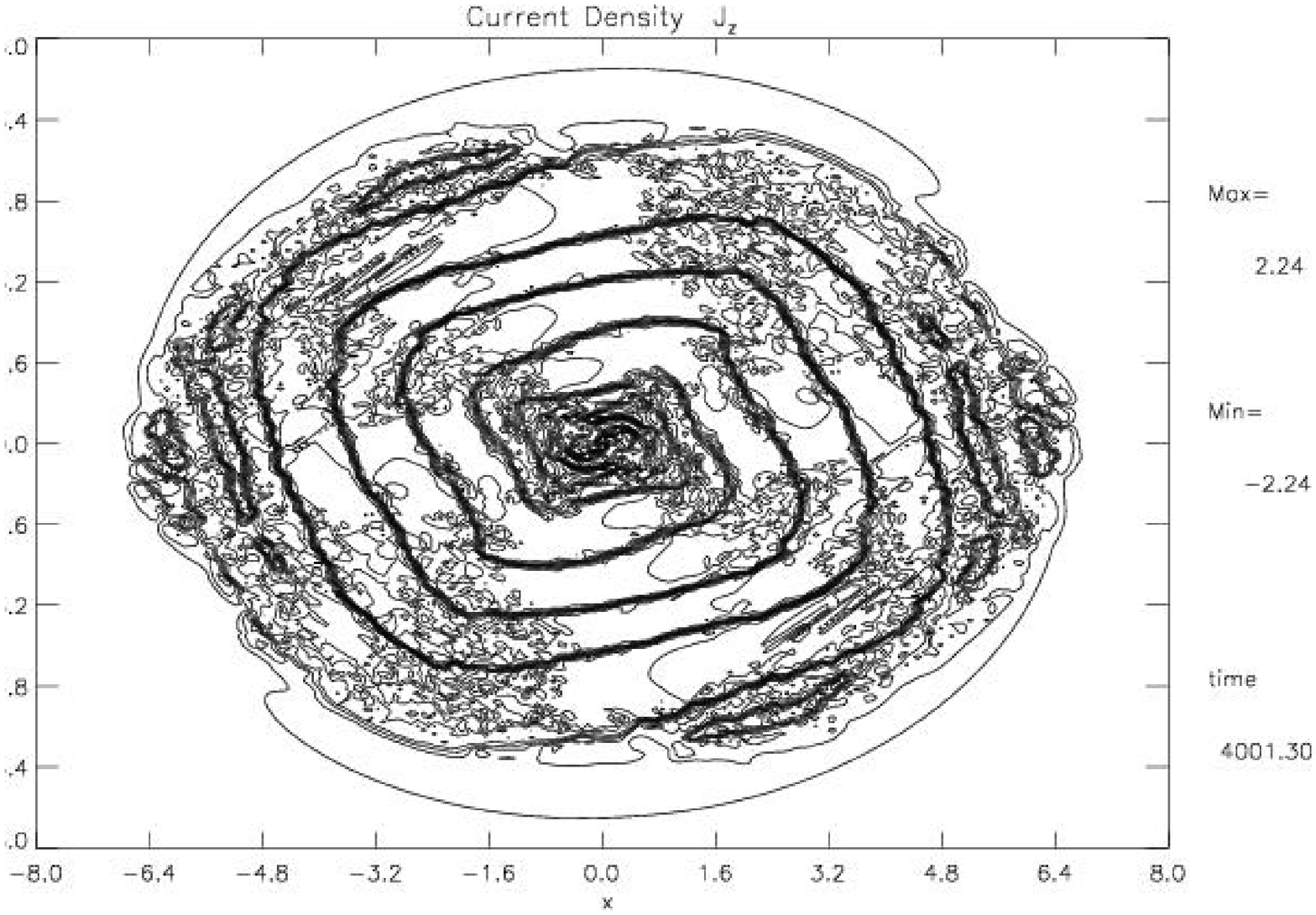}}
 \caption{Series of contour plots of the current density~\( j_{z} \) for the
 non-ideal case at different times}
 \label{fig:jzc60}
\end{figure}

\newpage

\begin{figure}[t]
  \center
  \includegraphics[angle=90,width=\picwid,keepaspectratio]%
	{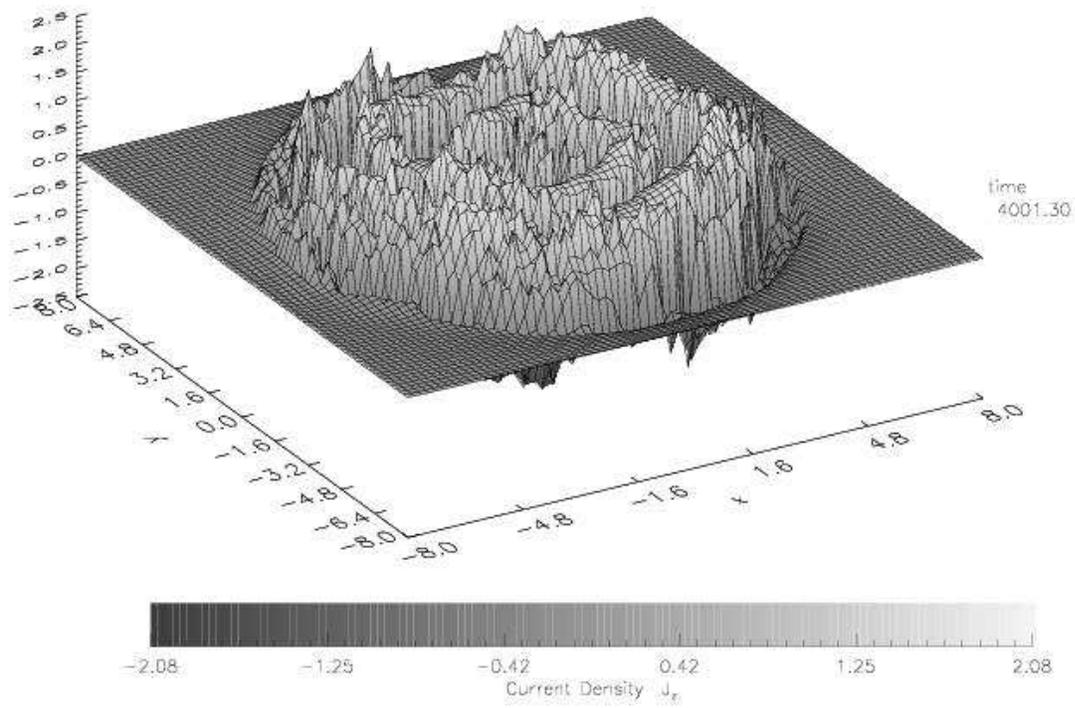}
  \caption{Surface plot of the current density~\( j_{z} \) at the end of the
  simulation (\(4001.3 \; \tau \ind{A} \))}
  \label{fig:sjzc68}
\end{figure}

\newpage

\begin{figure}[t]
  \center
  \includegraphics[angle=90,width=\picwid,keepaspectratio]%
	{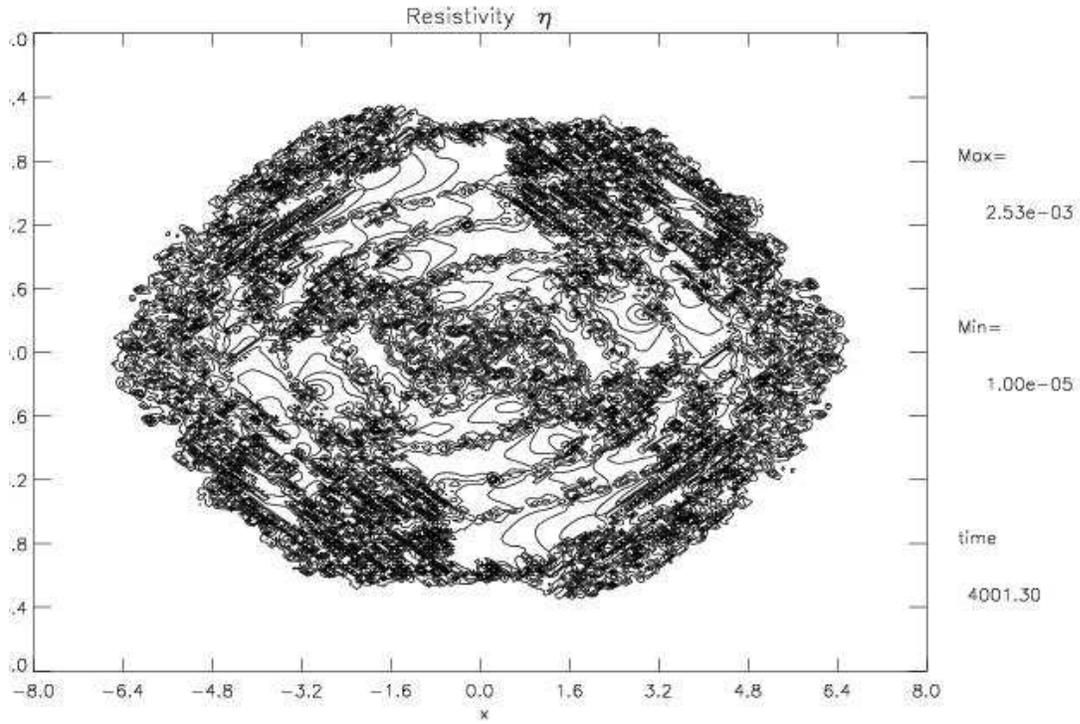}
  \caption{Contour plot of the resistivity~\( \eta \) at the end of the
  simulation (\(4001.3 \; \tau \ind{A} \))}
  \label{fig:resc68}
\end{figure}

\newpage

\begin{figure}[t]
  \center
  \includegraphics[angle=90,width=\picwid,keepaspectratio]%
	{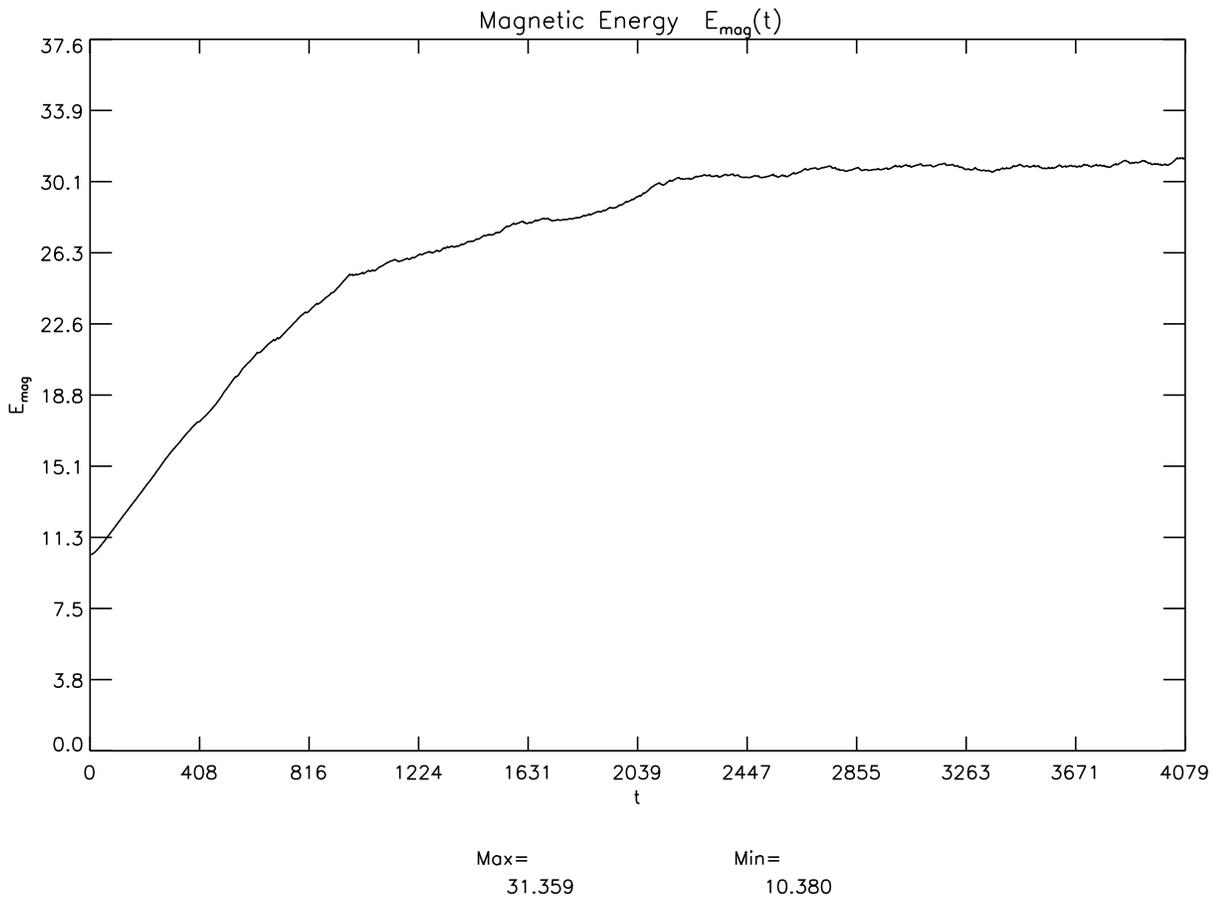}
  \caption{Temporal evolution of the magnetic energy of the system}
  \label{fig:magenergy}
\end{figure}

\newpage

\begin{figure}[t]
  \center
  \includegraphics[angle=90,width=\picwid,keepaspectratio]%
	{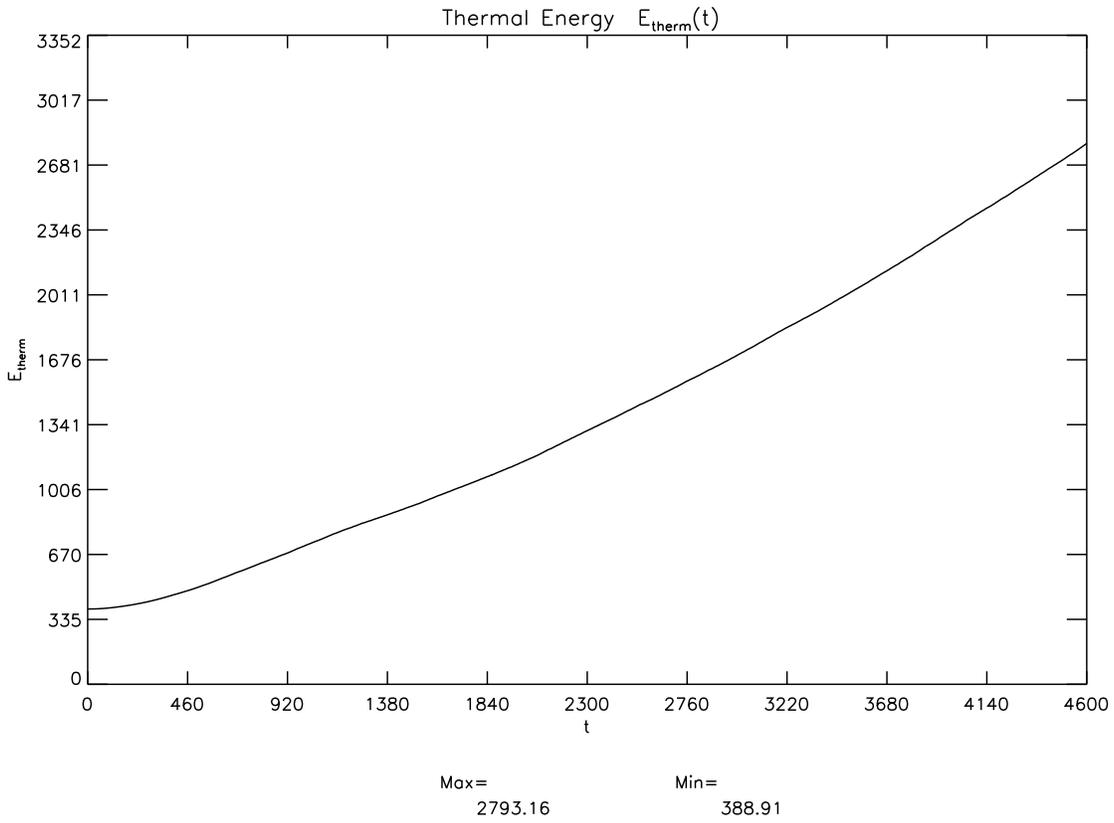}
  \caption{Temporal evolution of the thermal energy of the system}
  \label{fig:thermal}
\end{figure}

\newpage

\begin{figure}[t]
  \center
  \includegraphics[angle=90,width=\picwid,keepaspectratio]%
	{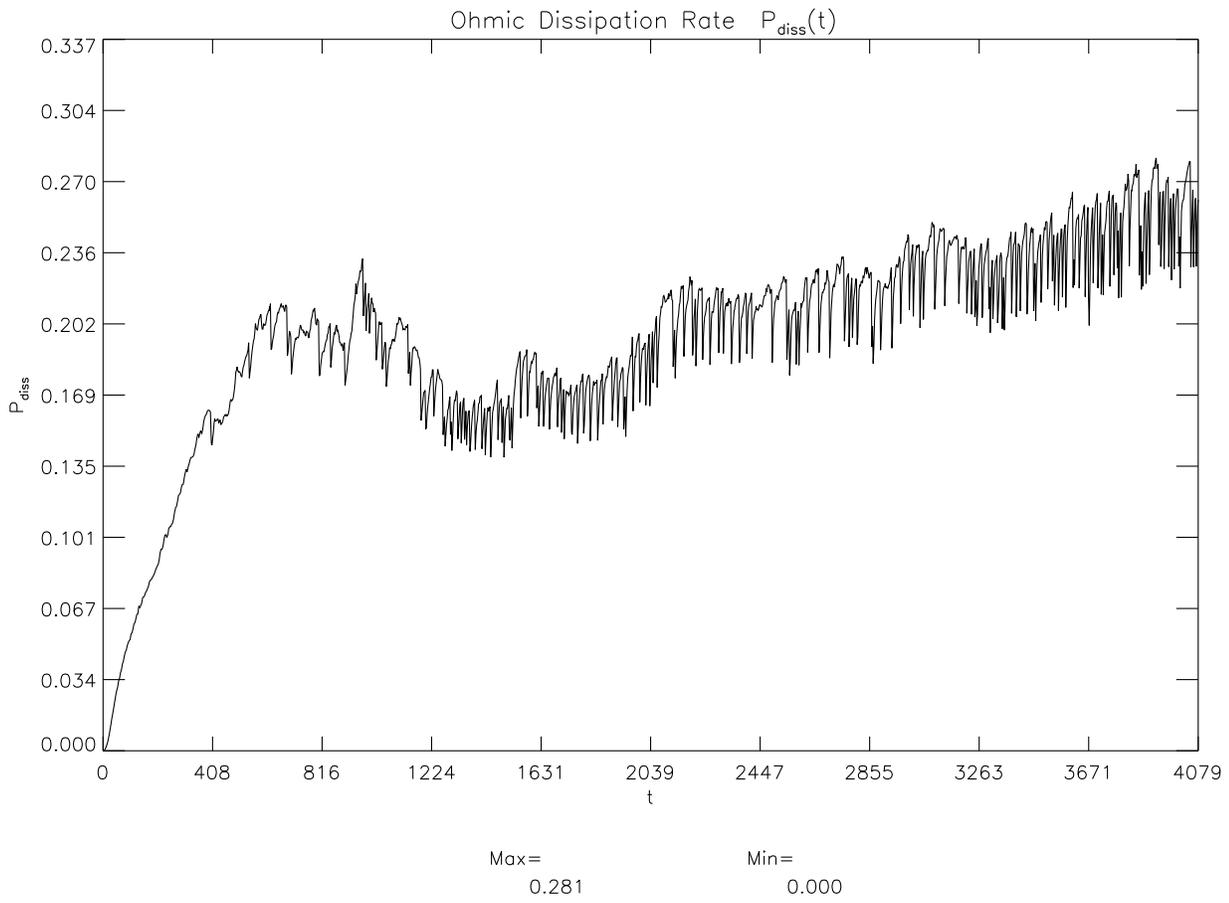}
  \caption{Temporal evolution of the Ohmic dissipation rate of the system}
  \label{fig:dissrate}
\end{figure}

\end{document}